\documentstyle[aps,prd,amssymb,amsbsy,eqsecnum,preprint,tighten,epsf]{revtex}

\begin{document}
\title{Data conditioning for gravitational wave detectors: A Kalman
  filter for regressing suspension violin modes}

\author{Lee Samuel Finn\thanks{Also Department of Physics, Department 
of Astronomy and Astrophysics; e-mail: LSF@Gravity.Phys.PSU.Edu}
and Soma Mukherjee\thanks{E-mail: Soma@Gravity.Phys.PSU.Edu}}
\address{Center for Gravitational Physics and Geometry, The
  Pennsylvania State University, University  Park, PA 16802, USA}

\maketitle

\begin{abstract}
  Interferometric gravitational wave detectors operate by sensing the
  differential light travel time between free test masses.
  Correspondingly, they are sensitive to anything that changes the
  physical distance between the test masses, including physical motion
  of the masses themselves. In ground-based detectors the test masses
  are suspended as pendula, in order that they be approximately
  ``free'' above the pendulumn frequency. Still, thermal or other
  excitations of the suspension wires' violin modes do impart a force
  on the masses that appears as a strong, albeit narrow-band,
  ``signal'' in the detectors wave-band. Gravitational waves, on the other
  hand, change the distance between the test masses without disturbing
  the suspensions. Consequently, violin modes can confound attempts to
  observe gravitational waves since ``signals'' that are correlated with
  a disturbance of the suspension violin modes are not likely due to a
  passing gravitational wave.
  
  Here we describe the design of a Kalman filter that determines the
  time-dependent vibrational state of a detector's suspension
  ``violin'' modes from time dependent observations of the detector
  output.  From the estimated state we can predict that component of
  the detector output due to suspension excitations, thermal or
  otherwise. The wire state can be examined for evidence of suspension
  disturbances that might, in the absence of such a diagnostic, be
  mistaken for gravitational wave signals. Additionally, from the wire
  state we can subtractively remove the contribution from suspension
  disturbances, thermal or otherwise, from the detector output,
  leaving a residual free from this instrumental artifact. We
  demonstrate the filter's effectiveness both through numerical
  simulations and application to real data taken on the LIGO~40~M
  prototype detector.
\end{abstract}

\section{Introduction}
 
Within the next several years five kilometer-scale interferometric
gravitational wave detectors (GEO~\cite{hough96a},
LIGO~\cite{abramovici92a} and VIRGO~\cite{bradaschia90a}) will begin
searching for astronomical sources of gravitational waves, something
the smaller TAMA~\cite{kuroda97a} prototype detector has already
begun.  The fundamental limitation on the sensitivity of these
terrestrial detectors is seismic noise at low frequencies, thermal
excitations of the mirror substrates and their suspensions at
intermediate frequencies, and laser shot noise at higher frequencies.
Here we focus on that part of the suspension noise associated with the
{\em violin modes\/} of the suspension wires. 

Gravitational waves interact with interferometric detectors by
changing the separation, and thus the round-trip light travel time,
between the mirrors in the detector arms. In order to be sensitive to
passing gravitational radiation, the mirrors in terrestrial
interferometeric detectors are suspended as pendula in order that they
be approximately inertially free above the pendulumn frequency.
Nevertheless, thermal and other excitations of the suspension wires
move the mirrors and lead to (narrow-band) data stream artifacts,
located at the violin-mode resonance frequencies of the suspension
wires.  These narrow-band features, which stand some 40--60~dB above
the shot noise background in the LIGO~I detector, are signals themselves
that, unless properly treated, can confound efforts to detect actual
gravitational wave signals.

In the waveband of the detectors now under development the transfer
function, which determines how the detector output at a particular
frequency is related to incident radiation at that frequency, is
constant. The time evolution of the signal arising from any incident
gravitational waves is thus determined solely by the waves themselves.
The time evolution of the signal arising from suspension violin mode
excitations, on the other hand, is strongly influenced by the the
dynamics of the modes. We make use of that distinction to design a
Kalman filter~\cite{kalman60a} that can, in real-time, detect violin
mode excitations --- thermal or otherwise --- as distinct from actual
gravitational wave signals. The measured excitation can be used to
subtractively remove the modes from the detector output, significantly
reducing the rms detector noise and simplifying the down-stream
analysis. Additionally, these measured excitations can be used to
diagnose the instrument's behavior.

Other techniques have been described for the identification and
removal of resonant line features in gravitational wave detector data.
In particular, Sintes and Schutz \cite{sintes98a,sintes99a} have
discussed methods suitable for addressing nearly periodic artifacts that
appear in a range of harmonics and whose phases, in each harmonic, are
linearly related.  These methods are not suitable for isolating the
violin modes, however: though a full spectrum of violin mode harmonics
do appear in the detector output, these modes are all independent and
independently excited. Allen and Ottewill \cite{allen98a} have applied
the multi-taper spectral method of Lees and Park\cite{lees95a} to
estimate the contributions of these resonant features to the power
spectrum. This non-parametric method does not and cannot provide a
time-domain estimate of the mode state or its instantaneous
contribution to the detector output and so cannot be used for
conditioning the data prior to analysis.

In section~\ref{sec:KFilter} we summarize the theory of Kalman
Filtering, laying the ground for its application to identifying
gravitational wave detector suspension violin modes.  The construction
of the Kalman filter for a particular problem requires a dynamical
model for the underlying physical process giving rise to the signal.
We discuss our model and how it comes to be represented as a set of
state-space difference equations on the physical observables. In
section~\ref{sec:investigation} we investigate the performance of the
filter using both simulated data sets, whose ``signal'' and ``noise''
content is known exactly, and experimental data taken at the LIGO~40~M
prototype detector during November 1994.\footnote{Preliminary results
  are presented in \cite{mukherjee99a}.} We summarize our conclusions in
section~\ref{sec:conclusions}.

\section{The Kalman Filter}\label{sec:KFilter}

\subsection{Introduction}

A Kalman filter is a mechanism for predicting the multi-dimensional
{\em state\/} of a dynamical system from a multi-dimensional {\em
observable.} The observable is assumed to be linearly related to the
system's state, which  is assumed to evolve linearly with time. 
The observable is assumed to be contaminated by an additive Gaussian
{\em measurement noise\/} of known co-variance.  The dynamical system
described by the state may be driven by a known input in addition to
Gaussian {\em process noise\/} of known co-variance.  

For example, the system may be a pendulum, the state may be the
pendulum bob's instantaneous position and momentum, and the observable
may be the bob's position. The process noise is the thermal
fluctuation in the wire and the measurement noise in this case may be
the photon shot noise background.

When the process and measurement noise are Normal ({\em i.e.,}
Gaussian) the Kalman filter gives the optimal estimate of the system's
state, in the least-squares sense.  When the noise is more complex
than Gaussian the Kalman filter gives the best estimate of the state
assuming that only the second order statistical characterization of
the noise is known.  Alternatively, when the evolution equations are
not linear or the observables are not related linearly to the state,
then the Kalman filter has a natural generalization that makes it the
optimal means of predicting the state from the observables in a
least-squares sense \cite{doob55a}.

The derivation of the Kalman Filter equations is straightforward and
instructive.\footnote{For a more detailed discussion of Kalman
  Filters, see \cite{kalman60a,brockwell87a,gersch93a,brown97a}.}  In
the particular application dealt with in this paper we assume that
there is no driving force (aside from the process noise) and that the
system state and observables are needed only at discrete times;
correspondingly, we limit our derivation to that case. Denote the
state of the system at sample time $k$ by $\bbox{\psi}[k]$ and the
observable at sample time $k$ by $\bbox{z}[k]$:
\begin{mathletters}
\begin{eqnarray}
\bbox{\psi}[k] &:=& \left(
\begin{array}{l}
\text{The state vector of the dynamical}\\ 
\text{system at sample $k$; a $M\times 1$ vector}
\end{array}
\right)\\
\bbox{z}[k] &:=& \left(
\begin{array}{l}
\text{The system observable at}\\ 
\text{sample $k$; a $N\times 1$ vector}
\end{array}
\right).
\end{eqnarray}

The state $\bbox{\psi}$ evolves from one sample to the next according
to the linear equation
\begin{equation}
\bbox{\psi}[k+1] = \bbox{A}\cdot\bbox{\psi}[k] + \bbox{w}[k],
\label{eq:evolve-psi}
\end{equation}
where $\bbox{A}$ is a linear operator and $\bbox{w}[k]$ is an instance
of {\em process noise:}
\begin{eqnarray}
\bbox{A} &:=& \left(
\begin{array}{l}
\text{Linear transformation that}\\
\text{takes $\bbox{\psi}[k]$ to $\bbox{\psi}[k+1]$;}\\
\text{a $M\times M$ matrix}
\end{array}
\right)\\
\bbox{w} &:=& \left(
\begin{array}{l}
\text{Process noise: a $M$-dimensional}\\
\text{Normally distributed random}\\
\text{variable; a $M\times1$ vector.}
\end{array}
\right)
\end{eqnarray}
The process noise is assumed to be zero mean and white. It is then
fully characterized by its co-variance matrix $\bbox{W}$:
\begin{equation}
\bbox{W} := E[{\bbox{w}\cdot\bbox{w}^{T}}],\label{eq:Wdef}
\end{equation}
\end{mathletters}
where $E[\cdot]$ represents an ensemble average across instances of
its argument. The process noise co-variance is a symmetric $M\times M$
constant matrix.

The observable $\bbox{z}$ at sample $k$ is related to the system state
by the linear equation
\begin{equation}
\bbox{z}[k] = \bbox{C}\cdot\bbox{\psi}[k] +
\bbox{v}[k],
\label{eq:measurement}
\end{equation}
where $\bbox{C}$ is a linear operator and $\bbox{v}[k]$ is an instance
of {\em measurement noise:}
\begin{eqnarray}
\bbox{C} &:=& \left(
\begin{array}{l}
\text{The linear transformation}\\
\text{that relates the state $\bbox{\psi}[k]$}\\
\text{to the observable $\bbox{z}[k]$; an}\\
\text{$N\times M$ matrix}
\end{array}
\right)\\
\bbox{v}[k] &:=& \left(
\begin{array}{l}
\text{Measurement noise: a $N$-dimensional}\\
\text{Normally distributed random variable;}\\
\text{a $N\times 1$ vector}
\end{array}
\right).
\end{eqnarray}
The measurement noise is assumed to be zero mean, white,\footnote{If
  the measurement noise is not white, it can be whitened by a
  suitable, linear redefinition of the observable.} and have
co-variance $\bbox{V}$:
\begin{equation}
\bbox{V} := E[{\bbox{v}\cdot\bbox{v}^{T}}],
\end{equation}
where $E[\cdot]$ represents an ensemble average across instances of
$\bbox{v}[k]$ of the measurement noise. The measurement noise
co-variance is a symmetric $N\times N$ matrix.

Finally, we assume that the measurement noise and process noise are
independent:
\begin{equation}
E[\bbox{v}\cdot\bbox{w}^T] = 0.
\end{equation}

\subsection{Predicting the state from observations}

Suppose that we have an estimate $\widehat{\bbox{\psi}}[k-1]$ of the
process state at sample $k-1$. Denote the error in this estimate by
$\widehat{\bbox{e}}[k-1]$:
\begin{equation}
  \widehat{\bbox{e}}[k-1] := \bbox{\psi}[k-1] -
  \widehat{\bbox{\psi}}[k-1].\label{eq:hat-e} 
\end{equation}
This error is a random variable whose second moments we denote
$\bbox{P}[k-1]$:
\begin{equation}
\widehat{\bbox{P}}[k-1] :=
E[\widehat{\bbox{e}}[k-1]\cdot\widehat{\bbox{e}}^{T}[k-1]].
\label{eq:hat-P} 
\end{equation}

From the estimated state at sample $k-1$, we can form an {\em a
  priori\/} estimate of the state at sample $k$ using the state
  transition operator $\bbox{A}$:
\begin{equation}
\widetilde{\bbox{\psi}}[k] := \bbox{A}\cdot\widehat{\bbox{\psi}}[k-1].
\label{eq:tilde-psi}
\end{equation}
Associated with 
$\widetilde{\bbox{\psi}}[k]$ is an error 
$\widetilde{\bbox{e}}[k]$ whose second moments we denote
$\widetilde{\bbox{P}}[k]$: 
\begin{mathletters}
\begin{eqnarray}
  \widetilde{\bbox{e}}[k] &:=& \bbox{\psi}[k] -
  \widetilde{\bbox{\psi}}[k].\label{eq:tilde-e} \\
\widetilde{\bbox{P}}[k] &:=&
E[\widetilde{\bbox{e}}[k]\cdot\widetilde{\bbox{e}}^{T}[k]]
  \label{eq:tilde-P}\\  
&=& \bbox{A}\cdot\widehat{P}[k-1]\cdot\bbox{A}^{T} + \bbox{W}
\label{eq:tilde-P-simp}. 
\end{eqnarray}
\end{mathletters}
The final equality follows from equations \ref{eq:tilde-e},
\ref{eq:evolve-psi} and \ref{eq:Wdef}. 

From $\widetilde{\psi}[k]$, we can make a {\em prediction\/}
$\widetilde{\bbox{z}}$ of the observable $\bbox{z}[k]$:
\begin{equation}
\widetilde{\bbox{z}}[k] = \bbox{C}\cdot\widetilde{\bbox{\psi}}[k].
\label{eq:tilde-z}
\end{equation}
Now compare the estimate $\widetilde{\bbox{z}}[k]$ to the actual
measurement $\bbox{z}[k]$. These will in general be different. The
Kalman Filter uses the difference between the the actual and estimated
{\em measurement\/} to generate a corrected estimate
$\widehat{\bbox{\psi}}$ of the {\em state:}
\begin{equation}
\widehat{\bbox{\psi}}[k] = \widetilde{\bbox{\psi}}[k] + 
\bbox{K}[k]\cdot\left(
\bbox{z}[k] - \widetilde{\bbox{z}}[k]
\right).\label{eq:hat-x}
\end{equation}
The linear operator $\bbox{K}[k]$ is the {\em Kalman gain,} which we
calculate in section \ref{sec:KGain}. The 
refined state estimate $\widehat{\bbox{\psi}}[k]$  has an associated
error $\widehat{\bbox{e}}[k]$ (cf.\ eq.\ \ref{eq:hat-e}) and covariance
$\widehat{\bbox{P}}[k]$ (cf.\ eq.\ \ref{eq:hat-P}):
\begin{equation}
\widehat{\bbox{P}} =
\left(\bbox{I}-\bbox{K}\cdot\bbox{C}\right)
\cdot\widetilde{\bbox{P}}\cdot
\left(\bbox{I}-\bbox{K}\cdot\bbox{C}\right)^{T},\label{eq:hat-P1}
\end{equation}
where $\bbox{I}$ is the identity matrix and we have used
\ref{eq:hat-e}, \ref{eq:tilde-e} and \ref{eq:hat-x} to represent
$\widehat{\bbox{P}}$ in terms of the unknown $\bbox{K}$ and the known
$\bbox{C}$ and $\widetilde{\bbox{P}}$, all at sample $k$.

\subsection{Calculating the Kalman Gain}\label{sec:KGain}

Our object is to find the Kalman gain $\bbox{K}$ that minimizes an
appropriate measure of the error $\widehat{\bbox{P}}$ in the state
estimate $\widehat{\psi}$ at each sample $k$.

Recall that $\widehat{\bbox{P}}$ is a co-variance matrix of the error
$\widehat{\bbox{e}}$; consequently, its eigenvectors describe linear
combination of the state vector elements whose errors are
uncorrelated. Correspondingly, the sum of the eigenvalues --- the
trace of $\widehat{\bbox{P}}$ --- measures the variance of the total
error in the state estimate.  Adopting $\text{tr}\,\widehat{\bbox{P}}$
as a measure of the error in the state estimate, we choose the Kalman
gain $\bbox{K}$ to minimize $\text{tr}\,\widehat{\bbox{P}}$:
\begin{eqnarray}
\bbox{K}[k] &=&
\widetilde{\bbox{P}}[k]\cdot\bbox{C}^T / 
\left(
\bbox{V} + \bbox{C}\cdot\widetilde{\bbox{P}}[k]\cdot\bbox{C}^T 
\right),
\label{eq:K}
\end{eqnarray}
where we have adopted the notation 
\begin{mathletters}
\begin{equation}
\bbox{Q} = \bbox{S}/\bbox{R}
\end{equation}
to denote $\bbox{Q}$ as a solution to the linear equation
\begin{equation}
\bbox{S} = \bbox{Q}\cdot\bbox{R}
\end{equation}
\end{mathletters}
for known $\bbox{R}$, $\bbox{S}$. 
Correspondingly, equation \ref{eq:hat-P1} for $\widehat{\bbox{P}}$ can
be simplified:
\begin{eqnarray}
\widehat{\bbox{P}} &=&
\left(\bbox{I}-\bbox{K}\cdot\bbox{C}\right)
\cdot\widetilde{\bbox{P}}\cdot
\left(\bbox{I}-\bbox{K}\cdot\bbox{C}\right)^{T}\nonumber\\
&=& \left(\bbox{I}-\bbox{K}\cdot\bbox{C}\right)
\cdot\widetilde{\bbox{P}}
-\left(\bbox{I} - \bbox{K}\cdot\bbox{C}\right)
\cdot\bbox{K}\cdot
 \left(\bbox{V} + \bbox{C}\cdot\bbox{K}\cdot\bbox{C}^{T}\right)
\bbox{K}^{T}+\bbox{K}\cdot\bbox{V}\cdot\bbox{K}\nonumber\\
&=&
\left(\bbox{I}-\bbox{K}\cdot\bbox{C}\right)
\cdot\widetilde{\bbox{P}}.\label{eq:hatP}
\end{eqnarray}

\subsection{Summary}\label{sec:recursive}

To summarize, we begin with an estimate of the state
$\widehat{\bbox{\psi}}$ and the associated error $\widehat{\bbox{P}}[k]$
at sample
$k-1$.  From these, we
\begin{enumerate}
\item Form an {\em a priori\/} estimate of the state
  $\widetilde{\bbox{\psi}}[k]$, the error associated with that
  estimate $\widetilde{\bbox{P}}[k]$, the predicted measurement
  $\widetilde{z}[k]$, and the Kalman gain $K[k]$, all at sample $k$
  (eqs.\ \ref{eq:tilde-psi}, \ref{eq:tilde-P-simp}, \ref{eq:tilde-z}, and
  \ref{eq:K});
\item Observe the system to determine $\bbox{z}[k]$;
\item Form corrected estimates of the state $\widehat{\psi}[k]$ and
  the estimated error $\widehat{\bbox{P}}[k]$ (cf.\ eq.\
  \ref{eq:hat-x}, \ref{eq:hatP}).
\end{enumerate}
We end as we begin, ready to make a new estimate of the state at
sample $k+1$. Once we are started, we can continue as long as
observations can be made. To begin, we need only supply an initial
estimate of the state and the error. These initial guesses are not
critical as the algorithm continually corrects the estimated state and
re-estimates the error with each iteration, converging on an optimum
(in the sense of minimum $\text{tr}\,\widehat{\bbox{P}}$) estimate of
the state.

The cost per iteration --- in either computation or memory --- is
constant.  Only the present estimates of the state and the associated
error need be retained from step to step: all of the history necessary
to estimate the future values of the state and the associated error is
encoded in these quantities.

\section{Application: Suspension violin modes}\label{sec:model}

\subsection{Introduction}

The cavity mirrors of terrestrial gravitational wave detector are
suspended as pendula in order that they be inertially free at
frequencies above the pendulum frequency.  The suspension wires are
under tension from the mirror mass and have a spectrum of ``violin''
modes.  Each violin mode is, like the pendulum mode, endowed with
$k_{B}T$ of energy from its contact with a thermal bath.  The
corresponding motion of the cavity mirrors leads to noise of thermal
origin in the detector output.  Other disturbances affect the
suspension, increasing the overall noise power in the violin modes.

\begin{figure}
  \epsfxsize=\columnwidth
  \epsffile{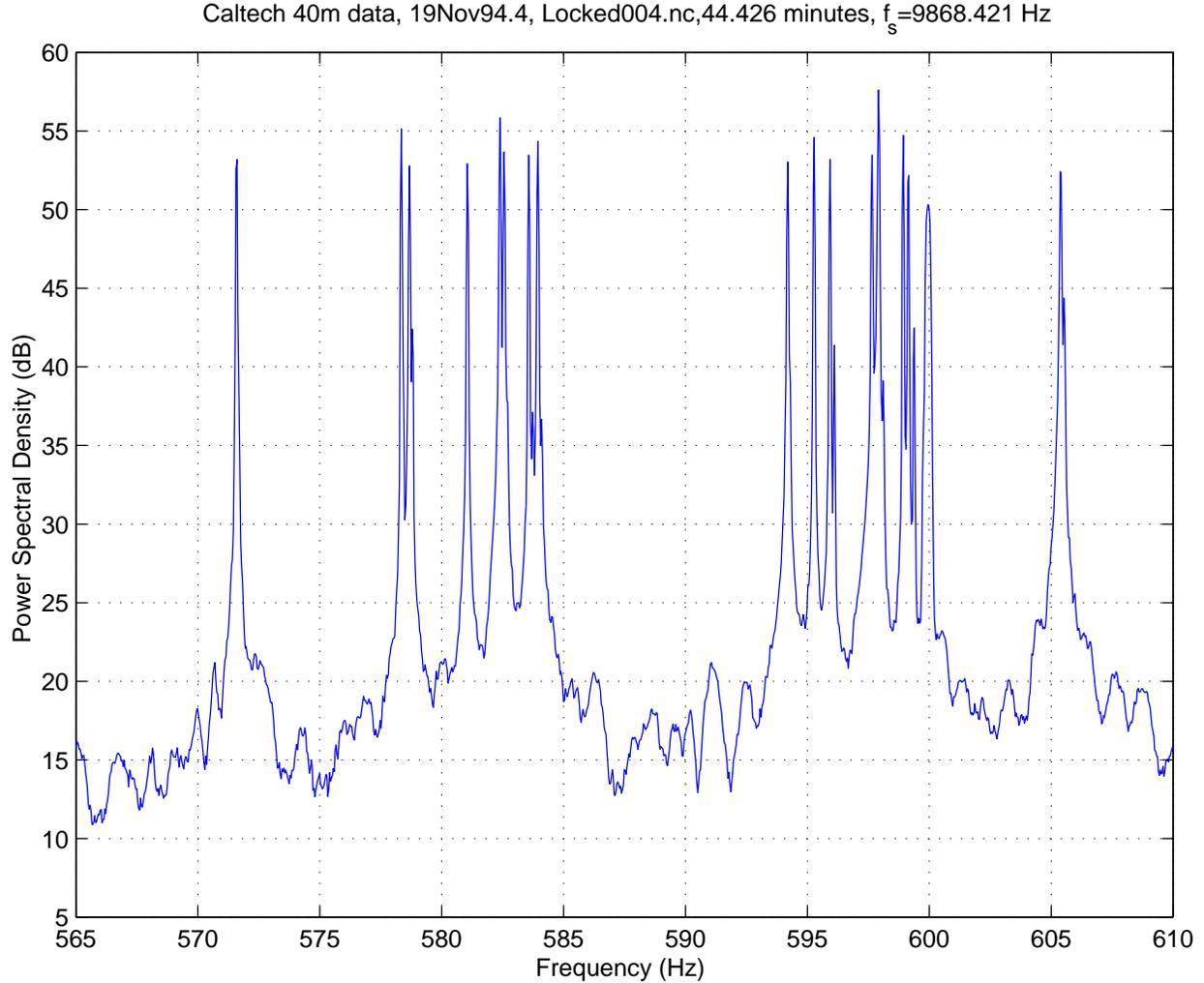}
  \caption{Typical power spectrum of the {\tt IFO\_DMRO\/} channel of
    the LIGO 40~M prototype, showing the violin modes between 571 and
    606 Hz. The units of the power spectrum are (ADC counts)${}^2$
    Hz${}^{-1}$.  Some lines are so closely spaced that they are not
    resolved on this figure. The line at 600~Hz is the ninth harmonic
    of the 60~Hz power-main.}\label{fig:violin}
\end{figure}

To minimize the impact of the thermal noise on the detector's
performance the suspensions are designed to have high $Q$.  Laser shot
noise dominates the detector noise power in a broad band about the
violin mode resonant frequencies: only the peaks of the stochastically
excited violin modes stand above this background.  The peaks are
strong, however: in the LIGO~40~M prototype they stand approximately
50~dB above the shot-noise floor and the contrast
in the LIGO~I detector will be approximately the same.  In the
LIGO~40~M prototype the fundamental violin mode resonant frequencies
are in the (570, 606)~Hz band with $Q$'s on order several times
$10^{4}$, while in the LIGO~I interferometer the fundamental violin
modes will have frequencies near 330~Hz and $Q$'s on order
$10^{5}$~\cite{gabypc}. Figure~\ref{fig:violin} shows the power
spectral density (PSD) of the differential-mode read-out
({\tt{}IFO\_DMRO}) of the LIGO~40~M prototype interferometer in the
neighborhood of the violin mode resonances for a typical stretch of
data taken during November 1994.  The units of the PSD are
ADC-counts${}^2$/Hz. The line at 600~Hz is the ninth harmonic of the
omnipresent 60~Hz power-main.
Note how the violin mode resonances stand-out as sharp peaks some
40~dB above an approximately white (over this bandwidth) background,
as described. 

The combination of violin modes in an approximately white background
maps naturally onto a model involving a stochastically driven
dynamical process observed in the presence of white measurement noise:
\begin{itemize}
\item Each violin mode corresponds to a distinct dynamical processes
driven by an independent noise source (thermal or otherwise);
    
\item The observable is the motion of the cavity mirror as reflected
in the gravity-wave channel of the interferometer;
    
\item The measurement noise is the laser shot-noise, and other
technical noises that affect the gravity-wave channel but do not drive
the violin modes.
\end{itemize}

In the remainder of this section we describe how we restrict attention
to the narrow band surrounding the violin mode, where the model
associated with the use of a Kalman filter is particularly apt, and
develop the state-space equations that describe the modes in that
band. 

\subsection{The Observations}\label{sec:obs}

The violin mode resonances are very weakly damped; so, the noise power
in the modes contributes significantly to the overall detector output
only in a very narrow band about each mode's resonant
frequency.  Correspondingly, to monitor one or more violin modes we
focus attention on a narrow band in the the neighborhood of the mode
resonant frequencies.  For definiteness, denote the gravitational wave
channel at full bandwidth (sample rate $f_{s}$) by $g$ and let the
band of interest, which includes the mode frequency $f_{0}$, range
from $f_{c}-\Delta f/2$ to $f_{c}+\Delta f/2$, for $\Delta f$ much
less than $f_{c}$.

To exploit the narrow bandwidth of the violin modes we form the complex
quantity 
\begin{equation}
    g'[k] = \exp\left(-2\pi i k{f_{c}\over f_{s}}\right) g[k]. 
    \label{eq:mix[k]}
\end{equation}
The spectrum of $g'$ in the band $[-\Delta f/2,\Delta f/2]$ is equal
to the spectrum of $g$ in the band $[f_{c}-\Delta f/2,f_{c}+\Delta
f/2]$.  Focus attention on just this narrow band by down-sampling $g'$
to the new sample rate $2\Delta f$ using, {\em e.g.,} a polyphase
technique \cite[Chapter 4]{strang96a}.\footnote{The down-sampling
  operation also involves the appropriate low-pass anti-alias
  filtering.} Denote the down-sampled $g'$ by $z$, which is
complex-valued.

From $z$, it is convenient to construct the real vector sequence 
$\bbox{z}[k]$:  
\begin{equation}
    \bbox{z}[k] = \left(
    \begin{array}{c}
        \Re(z)[k]\\
        \Im(z)[k]
    \end{array}
    \right).
\end{equation}
The vector-valued time series $\bbox{z}[k]$ plays the role of the 
observation in our Kalman filter implementation. 

\subsection{Dynamical Model}\label{sec:dynModel}

The gravitational wave channel $g$ is the sum of the violin modes
(whose behavior we wish to determine) and other noise sources.  In
constructing $\bbox{z}$ we focused attention on a narrow band about
the mode resonant frequencies.  Here we provide a model for
the mode behavior as reflected in $\bbox{z}$.

We first note that, near resonance, the details of the damping are
unimportant; consequently, we can model each violin mode as an
independent, viscously damped harmonic oscillator whose coordinate
$u$ is the amplitude of the mode present in the detector output.
The coordinate $u$ satisfies the differential equation
\begin{equation}
    \ddot{u} + {\omega_{0}\over Q}\dot{u} + \omega_{0}^{2}u = F(t),
\label{eq:dynamical}
\end{equation}
where $F$ is the (stochastic) driving force. The ``gravity-wave''
channel is the sum of $u$ and other noises (and signals).

We can relate $u$ to the contribution of the violin modes to
$\bbox{z}$ by mixing $u$ with the phase factor $\exp(-i \omega_{c}t)$
(where $\omega_{c}$ is equal to $2\pi f_{c}$):
\begin{equation}
    \psi = u\exp(-i\omega_{c}t). \label{eq:mix}
\end{equation}
The complex $\psi$ satisfies the differential equation
\begin{mathletters}
  \label{eq:LaplaceZ}
  \begin{equation}
    \ddot{\psi} + 
    \left({\omega_{0}\over Q}+2i\omega_{c}\right)\dot{\psi} + \left( 
      \omega_{0}^{2} - \omega_{c}^{2} + i{\omega_{c}\omega_{0} \over   
        Q}\right)\psi
    = F\exp(-i\omega_{c}t), 
  \end{equation}
  or, in the Laplace domain, 
  \begin{equation}
    \left(s-p_{{}+{}}\right)\left(s-p_{{}-{}}\right)\bar{\psi}(s) = 
    \bar{F}(s+i\omega_{c})
  \end{equation}
  where
  \begin{eqnarray}
    \bar{g}(s) &=&
    \int_{-\infty}^{\infty}dt\,\exp(-st)\,g(t)\\
    p_{\pm} &=& -{\omega_{0}\over2Q} - i\omega_{c} 
    \pm i\omega_{0}\sqrt{1-{1\over4Q^{2}}}.
  \end{eqnarray}
\end{mathletters}
(Note that in equations \ref{eq:LaplaceZ} we have assumed that $\psi$
is causal --- {\em i.e.,} it vanishes at sufficiently earlier times
--- and used a two-sided Laplace transform. Written in this way the
Laplace transform $\bar{\psi}$ of $\psi$ is related to its Fourier
transform $\tilde{\psi}$ through the substitution of $2\pi if$ for
$s$:
\begin{equation}
\tilde{g}(f) := \int_{-\infty}^\infty dt\,\exp(-2\pi i f t)\,g(t).
\end{equation}
(Note that we have adopted the engineering convention for the Fourier
transform.)  After down-sampling, the real and imaginary parts of
$\psi$ are the contributions of the violin modes to the real and
imaginary parts of $\bbox{z}$.

The two poles $p_{\pm}$ in the Laplace transform solution
$\bar{\psi}(s)$ correspond to the positive and negative frequency
resonances of the oscillator.  When we choose $\omega_{c}$ close to
$\omega_{0}$ we placed one of the resonances in $\psi$ near zero angular
frequency and the other far away: correspondingly, one of the poles
$p_{\pm}$ is near the origin and one is distant.  Letting $p_{+}$ be
the ``near'' pole and $p_{-}$ be the ``far'' pole we write
\begin{equation}
\left(s-p_{+}\right)\bar{\psi}(s) = {\bar{F}(s+i\omega_c)\over
  s-p_{-}}, \label{eq:a}
\end{equation}
or, for $|s|\ll|p_{{}-{}}|$,
\begin{equation}
\left(s-p_{+}\right)\bar{\psi}(s) =
-p_{-}^{-1}\bar{F}(s+i\omega_c). \label{eq:b} 
\end{equation}
The spectrum of $\psi$ arising from this second equation, which
differs from the first only by the source term, is identical to the
spectrum of $\psi$ arising from the first as long as $|\omega|$ is
much less than $|p_{-}|$. Since down-sampling restricts the bandwidth
of $\psi$ in just this way we adopt equation \ref{eq:b} as the
evolution equation for $\psi$.  Correspondingly, near zero angular
frequency $\psi$ satisfies
\begin{equation}
    \dot{\psi} - p_{+}\psi 
    = -p_{-}^{-1}F(t)e^{-i\omega_c t}.\label{eq:zeq}
\end{equation}

Now write, instead of the complex $\psi$, the real vector
$\bbox{\psi}$ whose components are the real and imaginary parts of
$\psi$:
\begin{equation}
    \bbox{\psi} = \left(
    \begin{array}{c}
        \Re(\psi)\\
        \Im(\psi)
    \end{array}
    \right). 
\end{equation}
The first order differential equation \ref{eq:zeq} describing complex
$\psi$ near zero frequency becomes a coupled set of first order
differential equations
\begin{mathletters}
  \begin{equation}
    \dot{\bbox{\psi}} = \bbox{A'}\cdot\bbox{\psi} +
    \bbox{F}
  \end{equation}
  where
  \begin{eqnarray}
    \bbox{A'} &=& \left(
      \begin{array}{cc}
\Re(p_{+})&-\Im(p_{+})\\
\Im(p_{+})&\Re(p_{+})
      \end{array}\right)\label{eq:A}\\
    \bbox{F} &=& F\left(\begin{array}{c}
        \Re(p_{-}^{-1}e^{-i\omega_c t})\\
        \Im(p_{-}^{-1}e^{-i\omega_c t})
      \end{array}\right).\label{eq:F}
  \end{eqnarray}
\end{mathletters}

Our observation is not of continuous $z(t)$ but of sampled
$\bbox{z}[k]$. Suppose that the final sampling rate is $\Delta f$ and
approximate the evolution of $\bbox{\psi}$ as free between samples,
with the accumulated effect of the stochastic force $\bbox{F}$ acting
at the sample times.  Then the sampled $\bbox{\psi}[k]$ satisfies
\begin{mathletters}
  \label{eq:state-space}
  \begin{equation}
    \bbox{\psi}[k] = \bbox{A}\cdot\bbox{\psi}[k-1]+\bbox{w}[k]
  \end{equation}
  where
  \begin{equation}
    \bbox{A} = \exp\left(\bbox{A}'\over\Delta f\right)
  \end{equation}
  and $\bbox{w}[k]$ is a random vector related to $\bbox{F}$.  The
  quantity $\bbox{\psi}[k]$ in this case is equal to the contribution
  of the violin mode to the observable $\bbox{z}[k]$, or
  \begin{equation}
    \bbox{z}[k] = \bbox{C}\cdot\bbox{\psi}[k] + \bbox{v}[k],
  \end{equation}
\end{mathletters}
where $\bbox{v}$ represents the (approximately) white background in
which the violin modes are immersed. Equations \ref{eq:state-space}
are the state-space form of the Kalman equations. To complete our
description of the suspension modes in this form it remains to
determine the properties of the process and measurement noises
$\bbox{w}$ and $\bbox{v}$.

The statistical properties of $\bbox{w}$ derive ultimately from
the statistical properties of the force $F$ (cf.~\ref{eq:dynamical}). 
In applying the Kalman filter we assume that $\bbox{w}$, and
consequently $F$, is white.  Assuming that modes are driven
principally by thermal forces we can use the Fluctuation-dissipation
theorem~\cite{callen51a} to find the power spectrum of the thermal
force driving the mode:
\begin{equation}
S_F(f) = 4k_B T\Re(Z)
\end{equation}
where $Z$ is the system's impedance. In our application,
\begin{equation}
Z = {-\omega^2+i\gamma\omega-\omega_0^2\over i\gamma\omega};
\end{equation}
so, $\Re(Z)$ is independent of frequency, the thermal force acting
on the modes is white, and the assumption made in applying the Kalman
filter to this problem is satisfied exactly. 

The quantity $\bbox{v}$ is the measurement noise: {\em i.e.,} the
noise within which the violin mode signal is embedded. In our
application the measurement noise is associated with the laser 
shot noise. In the current generation of large detectors the shot
noise has the spectrum 
\begin{equation}
S_{h}(f) = S_0\left[1+\left(f\over f_k\right)^2\right],
\end{equation}
with knee frequency $f_k$ approximately 100~Hz. the LIGO~40~M prototype
the mode fundamental frequencies are confined to an approximately
40~Hz bandwidth about 590~Hz; in the LIGO~I detector they will be in a
similarly narrow band about 350~Hz. As long as we are able to confine
attention to bandwidths small compared to the mode frequency the shot
noise in that band will be very nearly white; correspondingly,
$\bbox{v}$ will be white and have a covariance $\bbox{V}$ proportional
to the unity matrix.

To summarize, we construct the observable $\bbox{z}$ from the 
gravitational wave channel $g$ by mixing it with a local oscillator 
and down-sampling to the bandwidth $\Delta f$ about zero frequency. 
We model the sampled observable as arising from the equations
\begin{mathletters}
    \begin{eqnarray}
        \bbox{\psi}[k] &=& \bbox{A}\cdot\bbox{\psi}[k-1] + 
        \bbox{v}[k-1]\\
        \bbox{z}[k] &=& \bbox{C}\cdot\bbox{\psi}[k] + \bbox{w}[k]
    \end{eqnarray}
    where 
    \begin{eqnarray}
        \bbox{A} &=& \exp\left({\bbox{A}'\over\Delta f}\right)\\
        \bbox{A}' &=& \left(
        \begin{array}{cc}
          -{\omega_{0}/2Q}&\omega_{c}-\omega_{0}\sqrt{1-{1/4Q^{2}}}\\
          -\omega_{c}+\omega_{0}\sqrt{1-{1/4Q^{2}}}&-{\omega_{0}/2Q}
        \end{array}\right)
    \end{eqnarray}
\end{mathletters}

\subsection{Reconstruction}\label{sec:reconstruct}

In applying the Kalman filter we determine, at each sample, the state
estimate $\widehat{\bbox{\psi}}[k]$ from which we form an estimate
$\widehat{\bbox{z}}_v$ of the contribution of the violin mode to the
detector output:
\begin{equation}
\widehat{\bbox{z}}_v = \bbox{C}\cdot\widehat{\bbox{\psi}}. 
\label{eq:uprediction}
\end{equation}
The violin mode contribution to the observation is an instrumental
artifact that is unrelated to the incidence of gravitational waves on
the detector; correspondingly, we may wish to use the prediction to
subtractively remove the artifact from the data stream.\footnote{In
  section \ref{sec:characterization} we show that the Kalman
  filter estimate of the violin mode contribution to the observation
  is insensitive to the presence of a gravitational wave signal.}

The prediction $\widehat{\bbox{z}}_v$ for the violin mode contribution
is formed at reduced bandwidth and offset frequency.  From
$\widehat{\bbox{z}}_v$ we can construct the predicted contribution
violin mode contribution to the full bandwidth observation ($g[k]$,
cf.\ \ref{eq:mix[k]}) by reversing the steps taken to form $\bbox{z}$
from $g$ (cf.\ sec.\ \ref{sec:obs}). In particular, we first form the
complex scalar sequence
\begin{equation}
\widehat{z}_v = 
\left(\begin{array}{cc}1&i\end{array}\right)\cdot\widehat{\bbox{z}}_v
\end{equation}
and up-sample it to the full sample rate, forming $\widehat{G}'_v$. 
The mixing operation undertaken in equation \ref{eq:mix[k]} can then
be reversed,
\begin{equation}
\widehat{G}_v[k] = \widehat{G}'[k]\exp\left(2\pi i k{f_{c}\over
    f_{s}}\right).\label{eq:unmix}
\end{equation}
The result is still complex: it is missing the contributions from
negative frequencies that were discarded in the low-pass filtering
that accompanied the original down-sampling operation.  Since the
original observation was real, however, this contribution is just the
complex conjugate of the positive frequency contribution, which we
have; consequently, the estimated contribution of the violin mode to
the data stream $g[k]$ is
\begin{equation}
\widehat{g}_v[k] = 2\Re(\widehat{G}[k]).
\end{equation}
The residual difference 
\begin{equation}
g_r[k] = g[k] - \widehat{g}_v[k]
\end{equation}
is then free of violin mode artifact, to the accuracy of our
estimation procedure.

\subsection{Estimating the measurement and process
  noise}\label{sec:estnoise} 

To use the Kalman filter we must have a model for the process dynamics
($\bbox{A}$) and how the process contributes to the observable
($\bbox{C}$).  These we described for our particular application ---
the violin modes of an interferometer mirror suspensions (each mode an
independent process) --- in section \ref{sec:obs}.  Additionally, we
must have estimates for the process and measurement noise co-variance
matrices ($\bbox{W}$ and $\bbox{V}$).  Here we describe how we
estimate these from a short
segment of filter input, taken at the beginning of the input.

Consider the typical bandwidth of the observable $\bbox{z}$, shown in
figure \ref{fig:violin}.  The process contributes significantly over
only a small fraction of the bandwidth, where the resonance rises
above the measurement noise floor.  To estimate the measurement noise,
we form the PSD of $z$ and take the average power far from the
resonant peaks.  This average power $\overline{P}_{m}$ is the total
measurement noise ($\text{tr}\,\bbox{V}$).  We assume that there is no
cross-covariance between the noise in $\Re(z)$ and $\Im(z)$, in which
case
\begin{equation}
    \bbox{V} = {1\over2}\overline{P}_{m}\,
    \left(\begin{array}{cc}1&0\\0&1\end{array}\right).
\end{equation}

To estimate the process noise we focus on that part of the spectrum
near the resonant peak where the process dominates.  We assume, in our
model, that the process is driven by white noise.  From the Kalman
filter model, the power spectral density of the contribution to $z$ in
this bandwidth is given by (cf.\ equation \ref{eq:zeq})
\begin{equation}
    P_{z}(\omega) = {4Q^2\text{tr}\,\bbox{W}\over
    \omega_{0}^{2} + \left(
    2Q\omega_{c}-\omega_{0}\left(4Q^2-1\right)^{1/2}
    \right)^{2}}.
    \label{eq:West}
\end{equation}
With the exception of the total process noise ($\text{tr}\,\bbox{W}$)
the right hand side of equation \ref{eq:West} is known. The left hand
side we measure. To estimate $\text{tr}\bbox{W}$, we measure $P_{z}$
and sum it over the limited bandwidth dominated by the process.
Setting that equal to the same sum executed on the right hand side of
equation \ref{eq:West} we determine the proportionality constant
$\text{tr}\,\bbox{W}$. Assuming that the process noise is equally
distributed between the real and imaginary parts of $\bbox{z}$, with
no cross-variance, we thus determine the estimated process noise
covariance:
\begin{equation}
    \bbox{W} = {1\over2}\left(\text{tr}\,\bbox{W}\right)
    \left(\begin{array}{cc}1&0\\0&1\end{array}\right).
\end{equation}

\section{Characterization}\label{sec:characterization}

Before providing results from the application of our filter on
LIGO~40~M prototype data we characterize the filter's behavior on 
``synthetic'' or ``mock'' time series. These time series are
constructed so that the violin mode state, process noise, 
measurement noise, and signal content are known exactly.
By applying the filter to these data sets and investigating the
difference between its estimation of these quantities and their
actual, known values, we characterize the filter's performance. 

Four applications to synthetic data sets are described here:
\begin{enumerate}
\item A single isolated violin mode, driven by Gaussian white process
  noise and observed in the presence of Gaussian white measurement
  noise (cf.\~sec.~\ref{sec:case1});
\item Several closely spaced violin modes, driven by non-Gaussian
  white process noise and observed in the presence of non-Gaussian
  white measurement noise (cf.~sec.~\ref{sec:case2});
\item A single isolated violin mode, driven by Gaussian white process
  noise with a superposed impulsive excitation and observed in the
  presence of non-Gaussian white measurement noise
  (cf.~sec.~\ref{sec:case3});  
  and
\item A single isolated violin mode, driven by non-Gaussian white
  process noise and observed in the presence of non-Gaussian white
  measurement noise with a superposed impulsive excitation 
  (cf.~sec.~\ref{sec:case4}).
\end{enumerate}
The first case demonstrates the filter operating in the regime where
the measurement noise and process noise satisfy the assumptions made
in the derivation of the Kalman filter. The second case examines the
filter's ability to distinguish a violin mode from measurement noise,
and to distinguish between two different modes, when the noise is not
as simple as Gaussian and modes are closely spaced in frequency.  In
the third case we test the filter's ability to respond to impulsive
excitations of the suspension modes. In the final case we introduce
an impulse into the measurement noise and observe the filter's ability
to discriminate between it and the contribution from the violin mode.

As they appear in the detector data stream, violin modes are
characterized entirely by their resonant frequency $f_0$ and damping
constant $Q$.  Since we are dealing always with discrete-time signal
processing, the scale of the frequency does not matter: the only
relevant quantity (aside from $Q$) is the ratio of the resonant
frequency to the sampling frequency $f_s$.  Nevertheless, to
foreshadow our application of the filter to data taken at the LIGO~40~M
prototype, we assume that the sampling frequency is 9868.421~Hz, the
sampling rate of the LIGO~40~M prototype, and we always choose the
simulated violin modes to have resonant frequencies coincident with
the frequencies of modes identified in that instrument.  With this
convention, the frequency and damping constants of the simulated modes
used in the examples to follow are given in table \ref{tbl:SimParams}.

\subsection{Generating synthetic data sets}\label{sec:LineGen}

Before discussing the results of our investigations into the Kalman
filter's behavior we describe how we generated the random time series
used in those investigations. 

\begin{figure}
    \epsfxsize=\columnwidth
    \epsffile{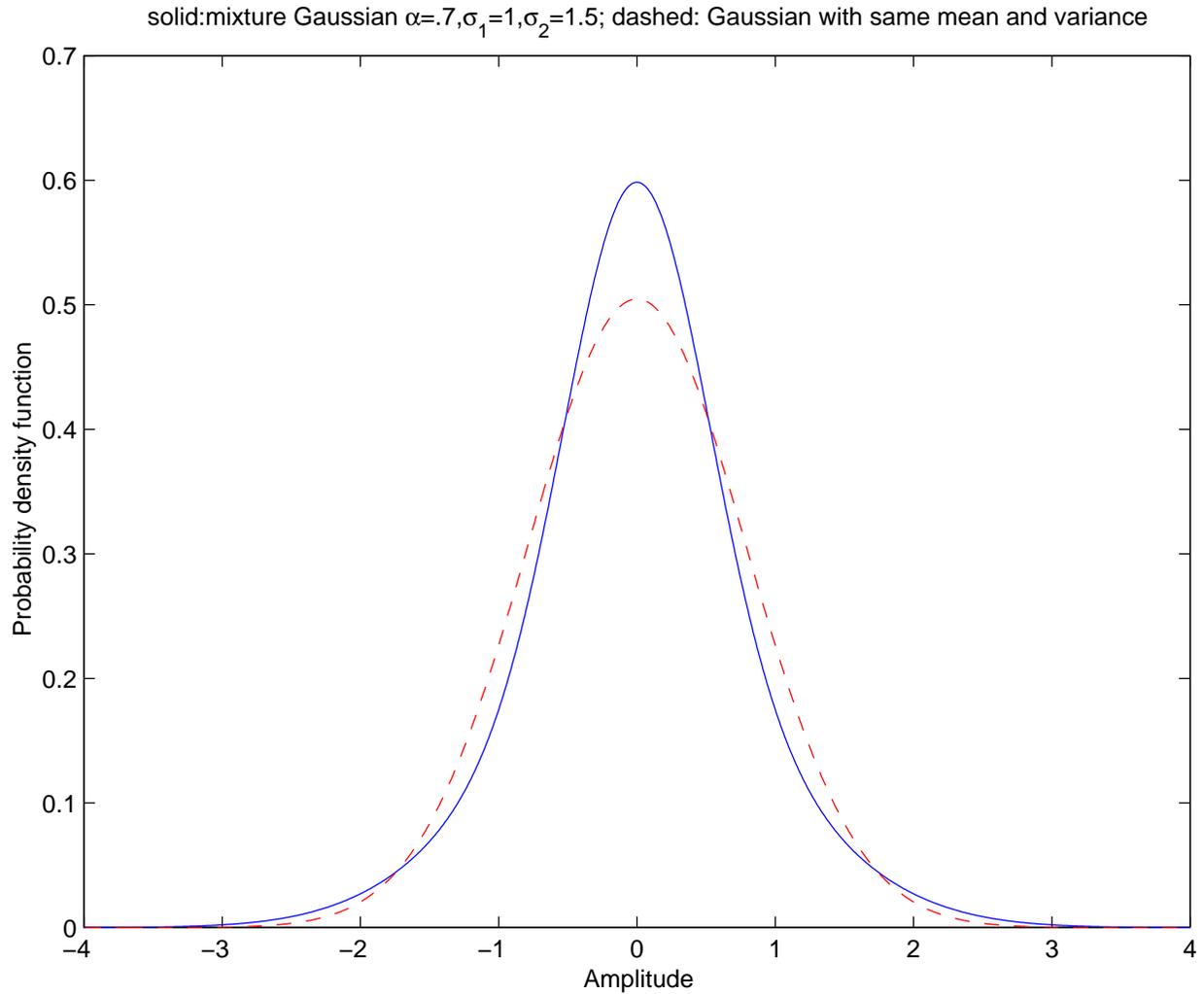}
    \caption{To test the effectiveness of the Kalman filter when the
      measurement and process noise is not Gaussian we used simulated
      data drawn from a strongly leptokurtic mixture Gaussian
      distribution. The solid curve shows the non-Gaussian noise
      distribution while the dashed curve shows, for comparison, a
      Gaussian distribution of the same mean and
      variance.}\label{fig:mgaussdist}
\end{figure}

In all cases the measurement noise is white: {\em i.e.,} its power
spectrum was constant over its bandwidth.  For Gaussian noise
measurement noise we used Matlab's \cite{matlab} {\tt randn\/} random
number generator. Non-Gaussian noise was modeled by a two-component
mixture-Gaussian model:
\begin{mathletters}
\begin{eqnarray}
P(x|\alpha,\mu_1,\mu_2,\sigma_1,\sigma_2) &=& 
\alpha N(\mu_1,\sigma_1) + \left(1-\alpha\right)N(\mu_2,\sigma_2)
\end{eqnarray}
\begin{eqnarray}
\begin{array}{rclcrcl}
\alpha   &=& 0.7,&\qquad&        & &   \\
\mu_1    &=& 0,  &      &\mu_2   &=& 0,\\
\sigma_1 &=& 1,  &      &\sigma_2&=& 1.5 .
\end{array}
\end{eqnarray}\label{eq:ex2ngparms}
\end{mathletters}
Here $N(\mu,\sigma)$ represents a Normal process with mean $\mu$ and
variance $\sigma^2$, the parameters $\alpha$, $\mu_k$ and $\sigma_k$
are fixed with $0\leq\alpha\leq1$.  Figure \ref{fig:mgaussdist} shows
the distribution of this mixture Gaussian process together with the
distribution of a Gaussian process with the same mean and variance. 
The mixture Gaussian distribution described by the parameters in
equation \ref{eq:ex2ngparms} is strongly leptokurtic ({\em i.e.,} it
has a significant excess of high amplitude events).

To generate the contribution to the observation from the process we 
begin with the differential equation modeling the process
(cf.\ \ref{eq:dynamical}):
\begin{equation}
    \ddot{r} + {\omega_{0}\over Q}\dot{r} + \omega_{0}^{2}r = F,
    \label{eq:rdifeq2}
\end{equation}
where $r$ is the position of the suspended mass. We make no
assumptions here about driving force $F$. The detector gravitational
wave channel consists of a superposition of measurement noise $v$ and
the oscillator coordinate $r$. The solution to this equation can be
expressed as an all-pole linear filter $H$ acting on the driving 
force $F$. In the Laplace domain the filter is given by 
\begin{mathletters}
\begin{eqnarray}
    \bar{r}(s) &=& \bar{H}(s)\bar{F}(s)\\
    \bar{H}(s) &=& {1\over s^{2} + s\omega_{0}/Q + 
\omega_{0}^{2}}\nonumber\\
    &=& {1\over\left(s-p_{+}\right)\left(s-p_{-}\right)}
    \label{eq:LaplaceH}
\end{eqnarray}
where
\begin{equation}
p_{\pm} = -\left(\omega_{0}/2Q\pm i\omega_{0}\sqrt{1-1/4Q^{2}}\right)
\end{equation}
\end{mathletters}
The frequency response of the filter at angular frequency $\omega$ is
equal to $\bar{H}(s)$ evaluated on the imaginary axis:
\begin{equation}
    \widetilde{H}(\omega) = \bar{H}(i\omega).\label{eq:contH}
\end{equation}

The two-pole filter $H$ operates in continuous time while the detector
output we wish to simulate is discretely sampled at a rate
$f_{s}$, which we assume is at least twice as great as the resonant
frequency $f_{0}$.  Using the analog filter $\bar{H}(s)$ as a
prototype, we design a digital filter with a similar response at
frequencies much less than the Nyquist frequency.

Suppose that the discretely sampled sequence $r[k]$, driven by the 
input $F[k]$, satisfies 
\begin{equation}
    u[k] = \sum_{j=0}^{N_{b}}b[j]F[k-j] - 
\sum_{j=1}^{N_{a}}a[j]u[k-j],
    \label{eq:causalLinFilt}
\end{equation}
where $a[j]$ and $b[j]$ are the constant filter coefficients. Define 
the 
$z$-transform of the sequence $y[k]$ by
\begin{equation}
    \check{y} \equiv \sum_{k=-\infty}^{\infty}y[k]z^{-k}.
\end{equation}
(The $z$-transform of a digital filter is analogous to the Laplace 
transform of a continuous filter.) 
In $z$-space, the linear filter described in equation 
\ref{eq:causalLinFilt} can be written
\begin{mathletters}
\begin{eqnarray}
    \check{r}(z) &=& 
    {\sum_{j=0}^{N_{b}}b[j]z^{-j}\over
    1+\sum_{j=1}^{N_{a}}a[j]z^{-j}}\check{F}(z)\\
    &=& \check{G}(z)\check{F}(z)
  \end{eqnarray}
\end{mathletters}

The frequency response of the digital filter $G$ is equal to 
$\check{G}(z)$ evaluated on the unit circle in the complex $z$-plane:
\begin{equation}
    \check{G}\left(e^{2\pi if/f_{s}}\right)
\end{equation}
where $f_{s}$ is the sample rate of the filtered sequence. If, by a
conformal transformation, we map the right half $s$-plane on to the
interior of the unit circle in the $z$-plane, then we will have mapped
the analog prototype $\bar{H}(s)$ on to the digital filter
$\check{G}(z)$:
\begin{equation}
    \check{G}(z) = \bar{H}(s)|_{s = 2\Omega(z-1)/(z+1)}, 
\end{equation}
where $\Omega$ is the so-called 
{\em warping constant.} This mapping will not be without distortion;
nevertheless, at frequencies much less than the Nyquist frequency the
distortion will be minimal. We can choose the warping constant
$\Omega$ to guarantee that the digital filter's response matches
exactly the analog filter's response at exactly one frequency:
\begin{equation}
    \check{G}(e^{-2\pi i f'/f_{s}}) = \bar{H}(2\pi i f')
\end{equation}
when 
\begin{equation}
    \Theta = {\pi f'\over\tan(\pi f'/f_{s})}.
\end{equation}
We choose the warping constant so that the response of the digital
filter matches the response of the analog filter at the resonant
frequency $f_0$. This completes the design of the digital filter
$\check{G}(z)$. Given a pseudo-random process $F[k]$ and the recursion
relationship of equation \ref{eq:causalLinFilt} we evaluate $r[k]$,
the simulated contribution of the violin mode to the detector output.
Finally, to obtain the simulated gravitational-wave channel output, we
add to $r[k]$ another pseudo-random process $n[k]$, which represents
the measurement noise.

In our simulations we used the {$\csc Matlab$} \cite{matlab} Normal
random number generator{\tt randn\/} wherever Normal deviates were
called for: {\em i.e.,} in forming either the measurement noise, the
process noise or the stochastic driving force $F$.

\subsection{Single mode: Normal, white process and measurement
  noise}\label{sec:case1} 

Using the techniques described above we generated several hundred
seconds of simulated data, consisting of a single mode with the
frequency and damping constants given in the first column of table
\ref{tbl:SimParams}. The process noise was white with variance of
2.38; the measurement noise was also white with variance 2.8.

\begin{figure}
  \epsfxsize=\columnwidth
\epsffile{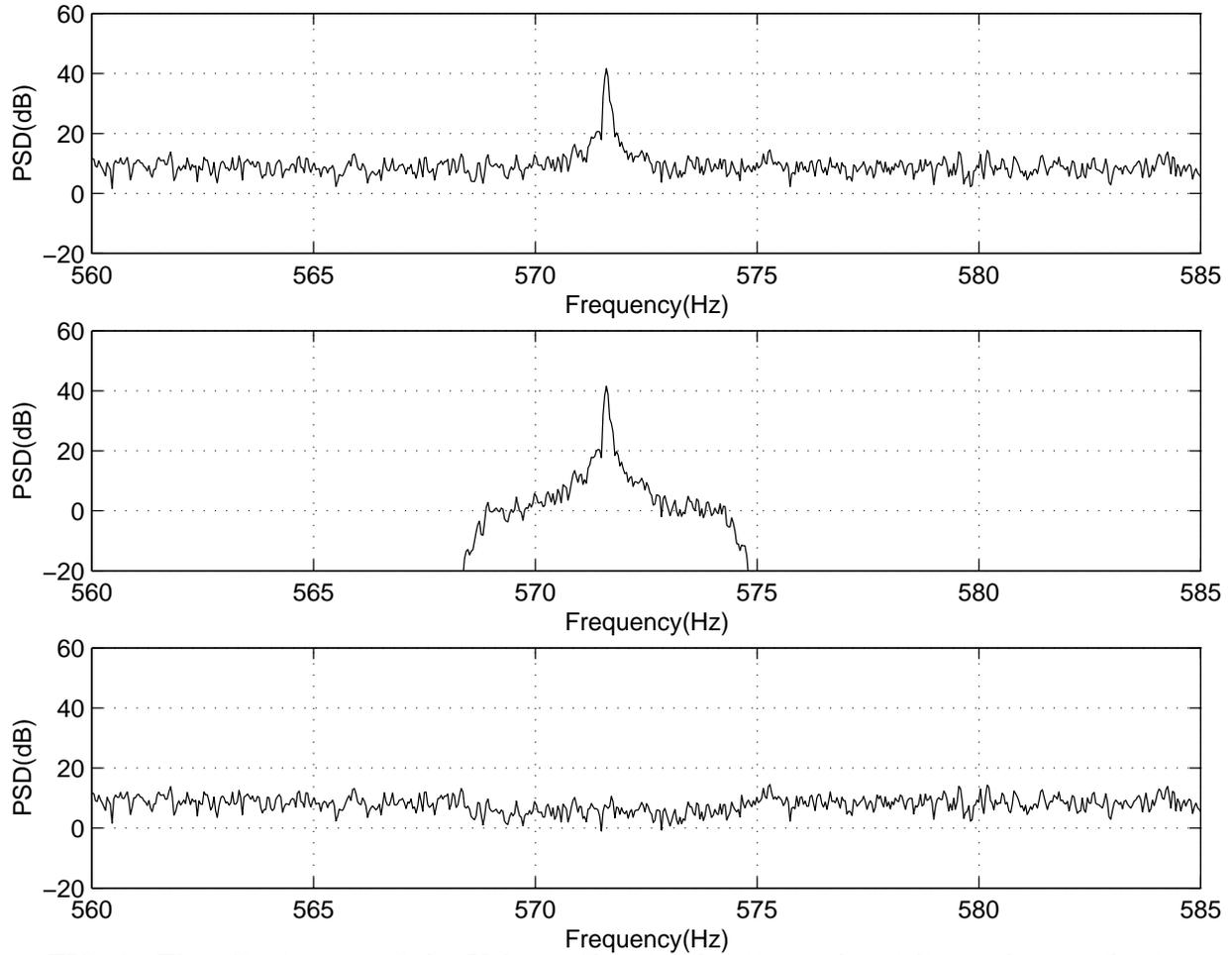}
  \caption{The effectiveness of the Kalman filter in identifying the
    violin mode contribution to the signal can be seen by comparing
    the power spectrum of the observed time series (top panel), the
    power spectrum of the Kalman estimated contribution of the mode to
    the observation (middle panel) and the power spectrum of observed
    time series less the filter's prediction of the violin mode's
    contribution to the observation (bottom panel). The bottom panel
    shows that the residual is, as expected, very nearly white.
    Kalman filter is applied to only a 6~Hz bandwidth about the line;
    this is the reason for the rapid fall-off of the predicted
    contribution of the mode to the total noise. The slight depression
    in the residual noise level in the signal band is discussed, along
    with other details, in
    section~\protect\ref{sec:case1}.}\label{fig:sim1a} 
\end{figure}

After generating the data set we applied the Kalman filter, focusing
on a 6~Hz bandwidth $\Delta f$ centered about the mode's resonant
frequency. Following the discussion in section \ref{sec:estnoise} we
estimated $\bbox{W}$ and $\bbox{V}$ using approximately 100~s of the
simulated data.\footnote{Since this data set is simulated the values
  of $\bbox{W}$ and $\bbox{V}$ are known {\em a priori;} however, in a
  real application they will not be known. We used the same procedure
  to estimate the filter as would be used in the field.}  We applied
the Kalman filter to the subsequent simulated data, using these values
of $\bbox{W}$, $\bbox{V}$, setting the initial state estimate
$\widehat{\bbox{\psi}}$ to zero and the initial estimated error
$\bbox{P}$ equal to $\bbox{W}\bbox{I}$ (where $\bbox{I}$ is the
identity matrix). Figure \ref{fig:sim1a} summarizes, in three panels,
the Kalman filter's effectiveness in estimating the mode state. The
upper panel shows the power spectrum of the simulated observations in
a wide band about the mode frequency. The middle panel shows the power
spectrum of the Kalman estimate of the mode contribution to the
observable.  Finally, the lower panel shows the power spectral density
of the difference between the actual data and the estimated
contribution from the mode.\footnote{While the power spectral density
  in a relatively wide band about the mode resonant frequency is
  shown, subtraction of the estimate affects only a 6~Hz bandwidth
  about the mixing frequency $f_c$ (cf.\ sec.\ \ref{sec:obs} and
  eqns.\ \ref{eq:mix}, \ref{eq:unmix}).}. Comparing these three panels
it is clear that the Kalman filter has correctly identified the modes
contribution to the observable.

\begin{figure}
  \epsfxsize=\columnwidth
  \epsffile{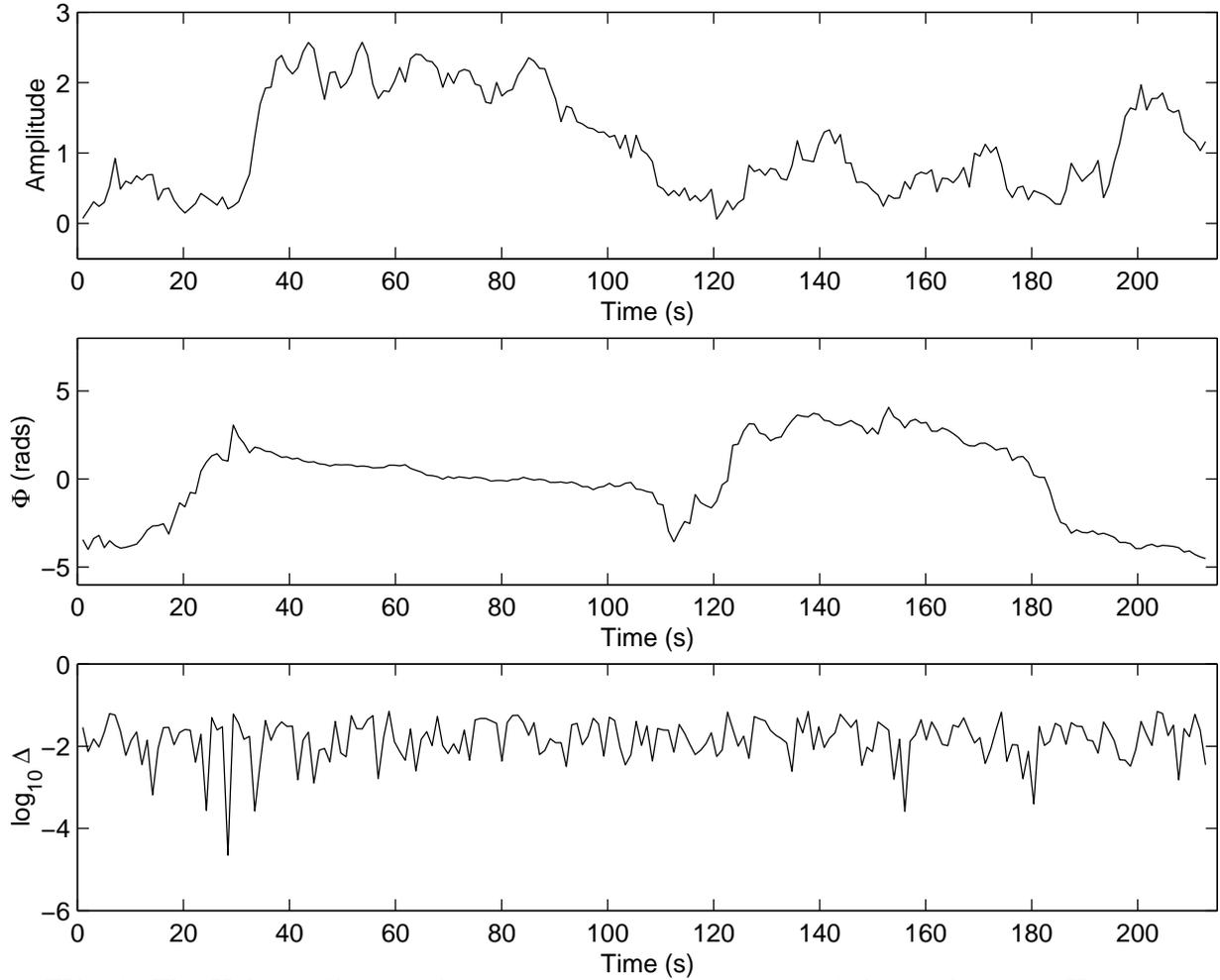}
  \caption{The Kalman filter produces instantaneous estimates of the
    mode state. Here we represent those estimates as an amplitude (top
    panel) and phase (relative to the resonant frequency angular
    frequency $2\pi f_0t$; center panel). In this
    example we know the actual state as a function of time. The bottom
    panel shows the instantaneous error in the state vector relative
    to its instantaneous magnitude. For more details see section
    \protect\ref{sec:case1}.}\label{fig:sim1c}
\end{figure}

Since the data is simulated we know exactly the measurement noise and
mode contributions to the observable. Correspondingly, we can evaluate
the actual error in the filter's estimate of the mode state. The mode
(estimated) state is described by the complex ($\widehat\psi[k]$)
$\psi[k]$, which we describe by its amplitude and phase
\begin{mathletters}
\label{eq:amp-phase}
\begin{eqnarray}
A &:=& |\psi|\label{eq:amp}\\
\tan\phi &:=& \Im(\psi)/\Re(\psi)\label{eq:phase}.
\end{eqnarray}
\end{mathletters}
The difference between the actual and estimated mode states we
describe by the quantity $\Delta$:
\begin{equation}
\Delta := {2|\psi[k] - \widehat{\psi}[k]|\over|\psi[k]+
  \widehat{\psi}[k]|}. 
\end{equation}
Figure \ref{fig:sim1c} shows the instantaneous estimated mode
amplitude, phase and error $\Delta$. The top panel shows the estimated
mode amplitude. The middle panel shows the estimate mode phase, less
$2\pi f_0 t$, where $f_0$ is the mode frequency. The bottom panel
shows the error in the estimated mode state.  Following an initial
epoch, during which the filter is ``locking-on'' to the mode state,
the relative error in the state estimate falls quickly to zero.

The middle and bottom panel of figure~\ref{fig:sim1a} show a slight
depression in the power spectral density of the residual in the band
that the Kalman filter operates. This slight depression is associated
with the error in the state estimate, which is shown in the bottom
panel in figure \ref{fig:sim1c} to be on order several percent. The
depression is uniform throughout the band, showing that the errors
introduced by the subtraction of the estimated mode contribution from
the data are white. For this reason it is unlikely that these errors
can be reduced by, for example, a ``better'' estimate of $\bbox{W}$,
$\bbox{V}$, $f_0$ or $Q$: changing any of these would lead to an error
that was not white, but reflected either an over- or under-estimate of
the power in the mode relative to the measurement error.

We expect that the continuously updated mode amplitude and phase
estimates as a function of time, either analyzed directly or displayed
in a strip chart of which the top two panels of figure~\ref{fig:sim1c}
may be considered a snapshot, will be particularly useful for
monitoring the state of the detector.

\subsection{Multiple modes; non-Gaussian white measurement and
  process noise}\label{sec:case2}

In deriving the Kalman filter we assumed that the process and
measurement noise were white; however, we never required that they be
Gaussian. Here we investigate the effectiveness of the filter when
these noises are non-Gaussian and when there are several modes
present, at least two of which are very close together in frequency.
This is a critical application, since in actual data there are
multiple closely spaced violin modes resulting from several suspension
wires of the mirrors, which cannot be estimated separately (cf.\ 
\ref{fig:violin}). The mode frequencies and damping constants for this
example are given in the second column of table \ref{tbl:SimParams}.

For isolated modes ({\em i.e.,} those whose resonant peaks do not
overlap with each other above the level of the measurement noise), one
can design and apply separately a Kalman filter for each mode. Closely
spaced modes cannot be treated in isolation; however, since the mode
dynamics are independent, the development of a Kalman filter that
treats $N$ modes jointly is straightforward:
\begin{itemize}
\item Referring to section \ref{sec:obs}, fix a carrier frequency
  $f_c$ and a bandwidth $\Delta f$ such that the modes of interest
  all have resonances in the band $[f_c-\Delta f/2,f_c+\Delta
  f/2]$. 
\item Considered separately, the dynamics of the mode $j$ ($1\leq
  j\leq N$) and the contribution $\bbox{z}_j$ of the mode to the
  measurement are described by the equations
\begin{mathletters}
\begin{eqnarray}
\bbox{\psi}_j[k] &=& \bbox{A}_j\cdot\bbox{\psi}_j[k-1] + 
\bbox{w}_j[k-1]\\
\bbox{z}_j[k] &=& \bbox{C}_j\cdot\bbox{\psi}_j[k].
\end{eqnarray}
\end{mathletters}
\item The Kalman equations describing the joint evolution of the $N$
  modes, and their contribution to the observation $\bbox{z}$ are then
\begin{mathletters}
  \begin{eqnarray}
    \bbox{\Psi}[k] &=& \bbox{A}\cdot\bbox{\Psi}[k-1] + \bbox{w}[k-1]\\
    \bbox{z}[k] &=& \bbox{C}\cdot\bbox{\Psi}[k] + \bbox{v}[k],
  \end{eqnarray}
  where
  \begin{eqnarray}
    \Psi &:=& 
    \left(\begin{array}{c}
        \bbox{\psi}_1\\
        \vdots\\
        \bbox{\psi}_N
      \end{array}\right),\\
    &=&
    \left(\begin{array}{c}
        \Re(\psi_1)\\
        \Im(\psi_1)\\
        \vdots\\
        \Re(\psi_N)\\
        \Im(\psi_N)
      \end{array}\right),\\
    \bbox{A} &:=& \left(\begin{array}{ccc}
        \bbox{A}_1&&\\
        &\ddots&\\
        &&\bbox{A}_N
      \end{array}\right),\\
    \bbox{W} &:=& \left(\begin{array}{ccc}
        \bbox{W}_1&&\\
        &\ddots&\\
        &&\bbox{W}_N
      \end{array}\right),\\
    \bbox{V} &:=& \left(\begin{array}{ccc}
        \bbox{V}_1&&\\
        &\ddots&\\
        &&\bbox{V}_N
      \end{array}\right),\\
    \bbox{C} &:=& 
    \left(\begin{array}{ccc}
        \bbox{C}_1&\cdots&\bbox{C}_N
      \end{array}\right)\\
    &=&\left(\begin{array}{ccccc}
        1&0&\cdots&1&0\\
        0&1&\cdots&0&1
      \end{array}\right),
  \end{eqnarray}
\end{mathletters}
and $\bbox{W}_j$ and $\bbox{V}_j$ are, as before, the process and
measurement noise.
\end{itemize}

In constructing $\bbox{W}$ we have approximated the individual modes
as being driven by independent process noise sources. Correlated
process noise leads to correlations in the states of the individual
modes. Even in the absence of an explicit correlation in the assumed
process noise covariance ({\em i.e.,} off-diagonal terms in
$\bbox{W}$), the filter will, if warranted by the observations,
correlate the state estimates. Lacking knowledge of the driving noise
(beyond contact with independent thermal baths), assuming that the
process noise is uncorrelated guarantees that there is no prejudice
that the mode states are correlated without constraining the state
estimates to be uncorrelated. 

\begin{figure}
  \epsfxsize=\columnwidth 
\epsffile{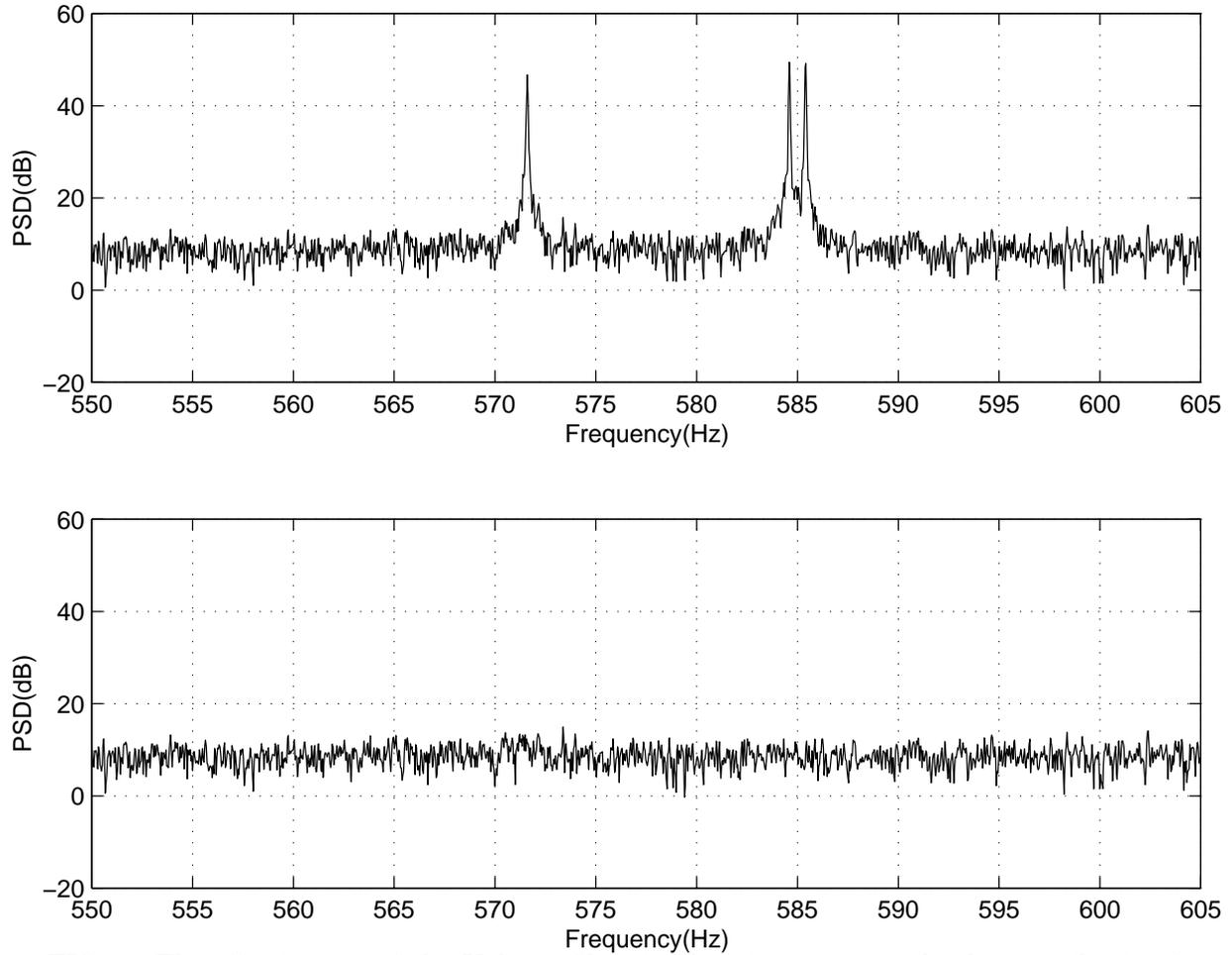}
  \caption{The effectiveness of the Kalman filter in identifying
    separately the contribution from two closely spaced violin modes
    can be seen by comparing the power spectrum of the observed time
    series (top panel) and the observed time series less the filter's
    prediction of the mode contributions. In this example the process
    noise and the measurement noise are both taken to be strongly
    leptokurtic. Despite the strong non-Gaussian character of the
    process and measurement noise the filter is able to identify all
    but $3\times10^{-6}$ of the power in the mode. For more details
    see section~\protect\ref{sec:case2}.}\label{fig:mixsub}
\end{figure}

To evaluate the effectiveness of the Kalman filter in treating
multiple modes driven by non-Normal noise, we applied a single Kalman
filter to this three mode system, choosing $f_c$ equal to 578.6~Hz and
$\Delta f$ equal to 18~Hz.  The top panel of figure \ref{fig:mixsub}
shows the power spectral density of the simulated gravity wave channel
in a 60~Hz bandwidth about the carrier frequency $f_c$, while the
bottom panel shows the power spectral density of the residual
difference between the simulated observation and the filter's
prediction of the contribution from the violin modes. The residual
broad feature that remains following the subtraction of the mode
estimate is due to the non-Gaussian nature of the process noise that
drives the mode. The filter has made a best-estimate of the mode
state, given only knowledge of the driving force variance. In this
example, however, the variance provides a very incomplete picture of
the process noise. Still, the filter has clearly identified
approximately 45~dB of the approximately 55~dB of the mode
contribution to the measurement.

\begin{figure}
    \epsfxsize=\columnwidth
    \epsffile{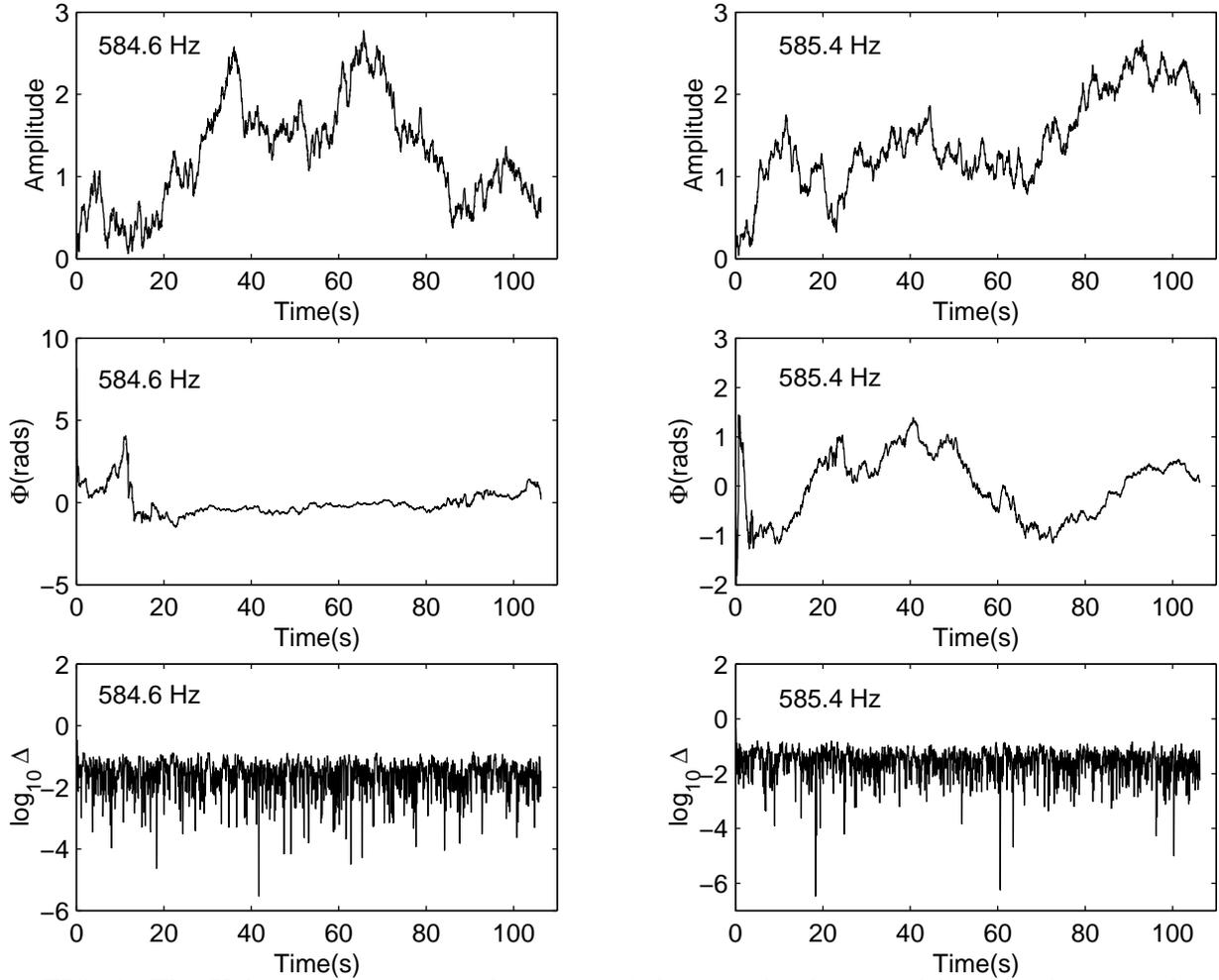}
    \caption{The Kalman estimates of the states of the 
      two closely spaced lines in the example of
      section~\protect\ref{sec:case2}. The right panels show the
      estimated amplitude (top), phase (less $2\pi f_0t$; middle) and
      fractional error $\Delta$ (bottom) for the $f_0 = 585.4$~Hz
      line; the left panels show the same quantities for the $f_0 =
      584.6$~Hz line. As is apparent from the errors the filter has
      successfully discriminated between the two lines.}\label{fig:mix2lines}
\end{figure}

The state $\bbox{\Psi}$ is the direct sum of the state vector
estimates $\bbox{\psi}_j$ of the individual modes; consequently, the
filter identifies not just the total contribution of each mode to the
observation, but the state of each mode separately.  The four panels
of figure~\ref{fig:mix2lines} show the estimated amplitude and phase
(less $2 \pi f_j t$ for each mode) of each of the two overlapping
modes at 584.6~Hz ($f_2$) and 585.4~Hz ($f_3$). In this example the
two modes are known to be uncorrelated; consequently, we expect that
the state estimates should also be uncorrelated, which we have found
to be the case. 

\subsection{Effect of a transient excitation of a 
mode}\label{sec:case3}

\begin{figure}
  \epsfxsize=\columnwidth
  \epsffile{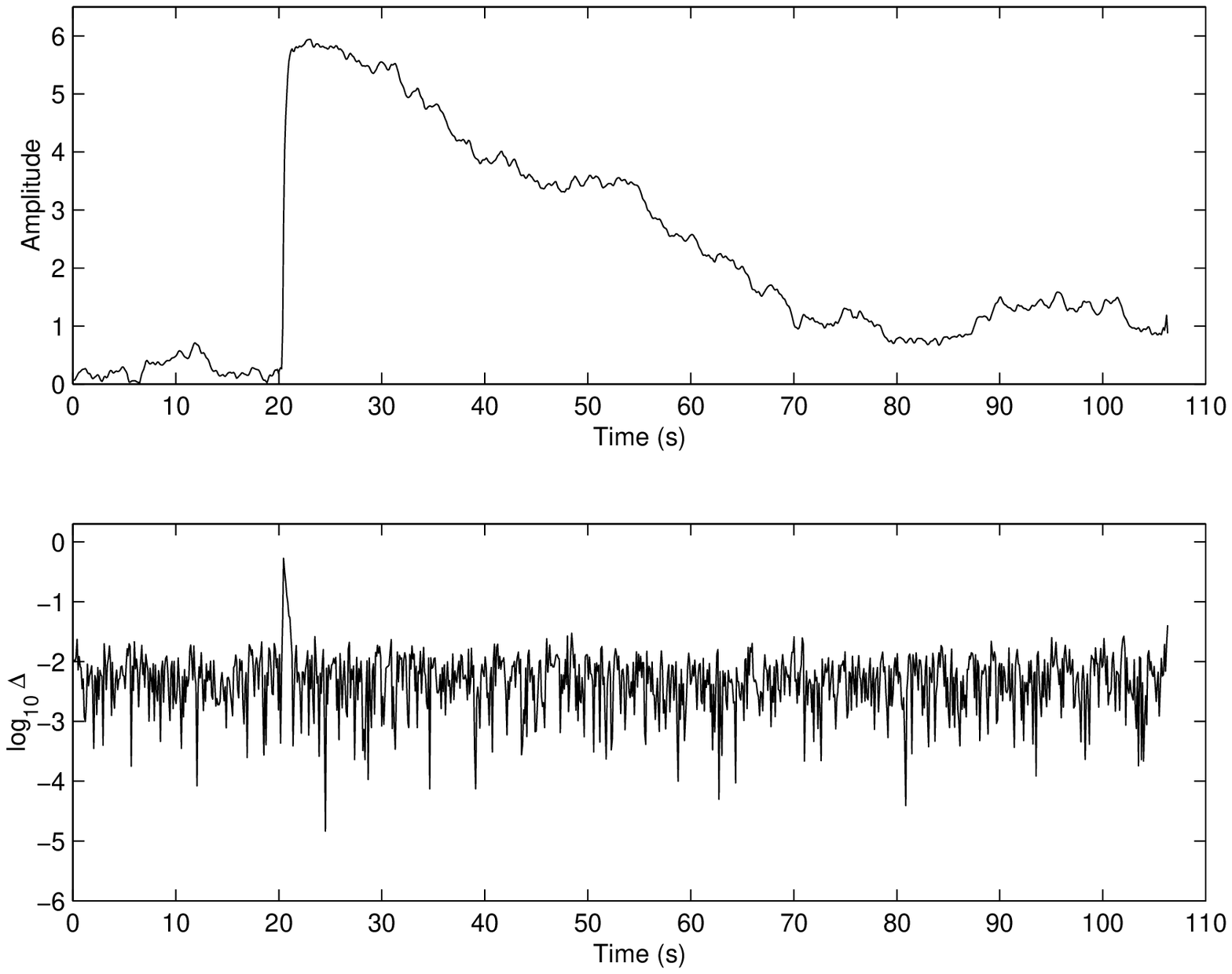}
  \caption{Can the Kalman filter track a mode that undergoes an
    impulsive excitation? At 20s into the simulation the mode was
    excited by a large force. The top panel shows the amplitude of the
    estimated mode state; the bottom panel shows the fractional error
    $\Delta$ (cf.\ eq.\ \ref{eq:Delta}) in the mode state estimate.
    The estimate is momentarily upset; however, the filter quickly
    re-acquires the actual mode state. For more detail see section
    \protect\ref{sec:case3}.}\label{fig:bang1}
\end{figure}

To evaluate the effectiveness of the Kalman filter in estimating the
filter state in the presence of a transient excitation, we have
simulated a single mode observed in the presence of the same
mixture-Gaussian measurement noise described above (cf.\ equation
\ref{eq:ex2ngparms}). The process noise is taken to be Gaussian with
variance 2.38. At 20 seconds into the simulation we add to the process
noise a single pulse of magnitude $35\sigma$, where $\sigma$ is the
rms amplitude of the process noise.  The top panel of figure
\ref{fig:bang1} shows the estimated amplitude of the mode while the
bottom panel shows error $\Delta$ in the mode state estimate:
\begin{equation}
\Delta := {2|\psi[k]-\widehat{\psi}[k]|\over
|\psi[k]+\widehat{\psi}[k]|}. \label{eq:Delta}
\end{equation}
After the filter's start-up transient the error in the estimated state
is small until the transient excitation. There is, not surprisingly, a
sudden increase in the error at the moment of the excitation; however,
the filter quickly adjusts to the new state. 

\subsection{Effect of transients in the measurement
  noise}\label{sec:case4} 

Sudden excitations in the observed time series may arise from sources
other than excitations of the violin modes.  To evaluate the
effectiveness of the Kalman filter in discriminating between
excitations of the mode and measurement noise transients we
simulate a single line, driven by Gaussian noise. This mode is
embedded in the same mixture-Gaussian measurement noise as described
in section \ref{sec:case2}, but with a pulse in the measurement noise
14 seconds into the simulation. The pulse lasts for a single sample
and has an amplitude of $35\sigma$, where $\sigma$ rms amplitude of
the measurement noise.

\begin{figure}
  \epsfxsize=\columnwidth
  \epsffile{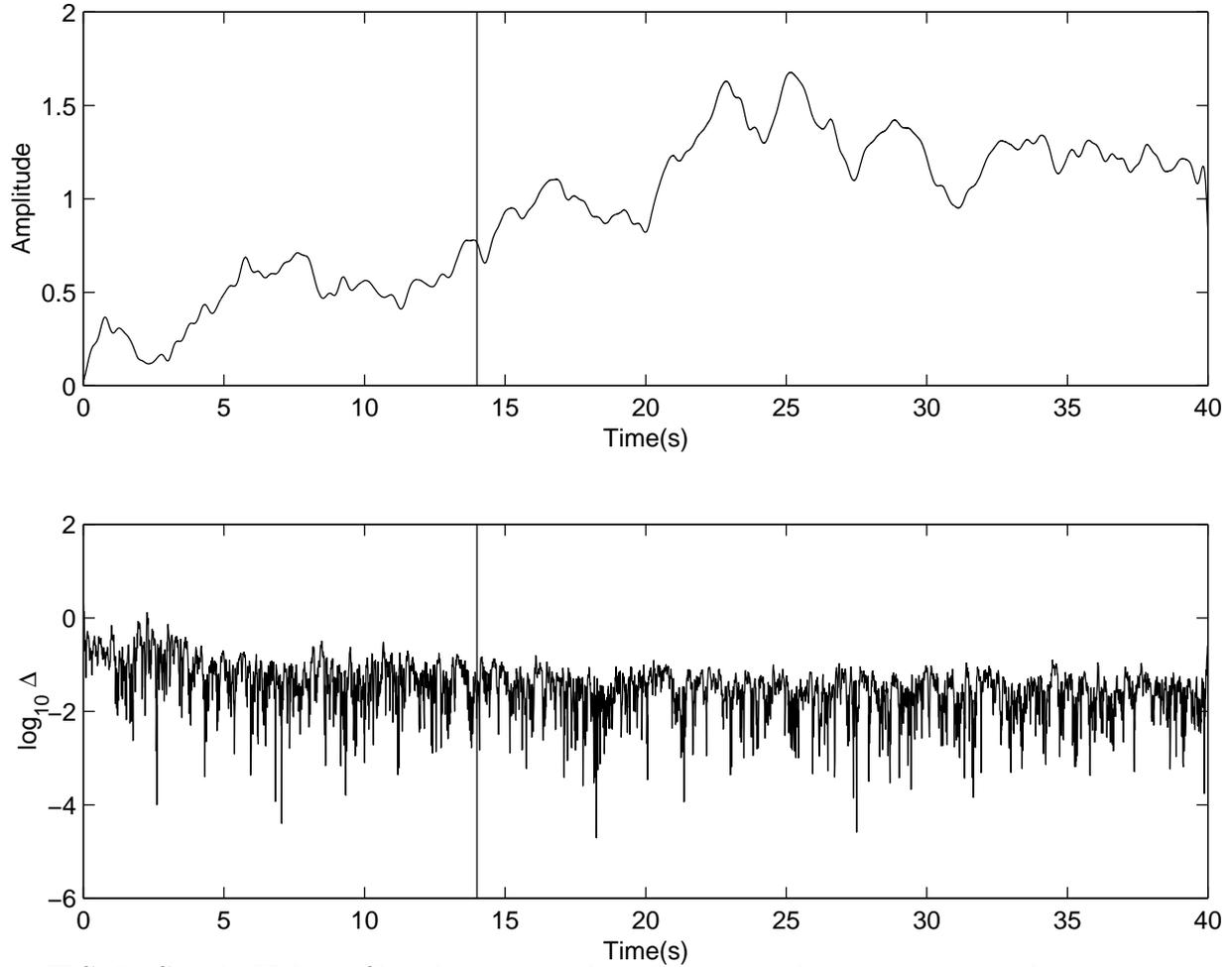}
  \caption{Can the Kalman filter discriminate between an impulsive
    excitation in the measurement noise and the contribution of the
    mode? A mode, stochastically driven by Gaussian process noise, is
    embedded in a mixture-Gaussian measurement noise. At 14s a large
    transient is added to the process noise. The top panel shows the
    Kalman estimate of the mode state amplitude throughout the
    simulation. The bottom panel shows the fractional error $\Delta$
    (cf.\ eq.\ \ref{eq:Delta}) in the estimated state. There is no
    evidence that the filter is sensitive to the transient measurement
    noise.  For more detail see section
    \protect\ref{sec:case4}.}\label{fig:bang2}
\end{figure}

The lower panel of figure \ref{fig:bang2} shows $\Delta$ (cf.\ 
\ref{eq:Delta}), measuring the magnitude of the difference in the
estimated and actual mode state; the upper panel shows the estimated
state itself. The error in the mode state estimate always remains
quite small, showing that the filter discriminates between excitations
of the modes and excitations in the measurement noise.

\subsection{Summary}

The Kalman filter described here is able to correctly determine
instantaneous state of the violin mode suspensions. This is true
whether the mode is driven by, or observed in the presence of,
Gaussian or non-Gaussian noise. The filter does not require that the
measurement or process noise be stationary: it correctly tracks the
mode after it has been impulsively excited and can distinguish between
excitations of the mode and transients in the measurement noise. 

\section{Violin modes in LIGO~40~M prototype
  data}\label{sec:investigation}

In this section we discuss the application of the Kalman filter to
data taken at the LIGO~40~M prototype detector during November 1994.
The general parameters of this run are described elsewhere
\cite{gillespie95a}.  The particular results presented here all refer
to the particular locked segment beginning at GPS time 469357019~s on
19 November 1994 and lasting for 2665.54496~s.

\subsection{General character of the resonant
  features}\label{sec:genchar} 

At the time of the November 1994 data run each of the four cavity
mirrors in the 40~M prototype was suspended by four cylindrical wires.
Setting aside the coupling of the wires through their attachment to
the suspended mirror there are two fundamental modes associated with
each wire: one with motion nominally along the optical axis and one
with motion perpendicular to that axis.  Owing to imperfections in the
wires, the attachments of the wires to the test masses, and the
coupling of the modes through the motion of the mass all the modes
will have some component along the optical axis.  We thus expect 32
fundamental violin modes.

A high resolution power spectrum focused on the bandwidth (570~Hz,
606~Hz) and taken of the data from a single locked-segment ({\em
  i.e.,} a single epoch during which the interferometer operated
without interruption as a gravitational wave detector) of 44~m
duration shows 28 distinct resonant features.  One of these is clearly
identified as the ninth harmonic of the 60~Hz power-main, leaving 27
different modes whose frequencies and quality factors are listed in
table \ref{tbl:lines}.
 
\begin{figure}
\epsfxsize=\columnwidth 
\epsffile{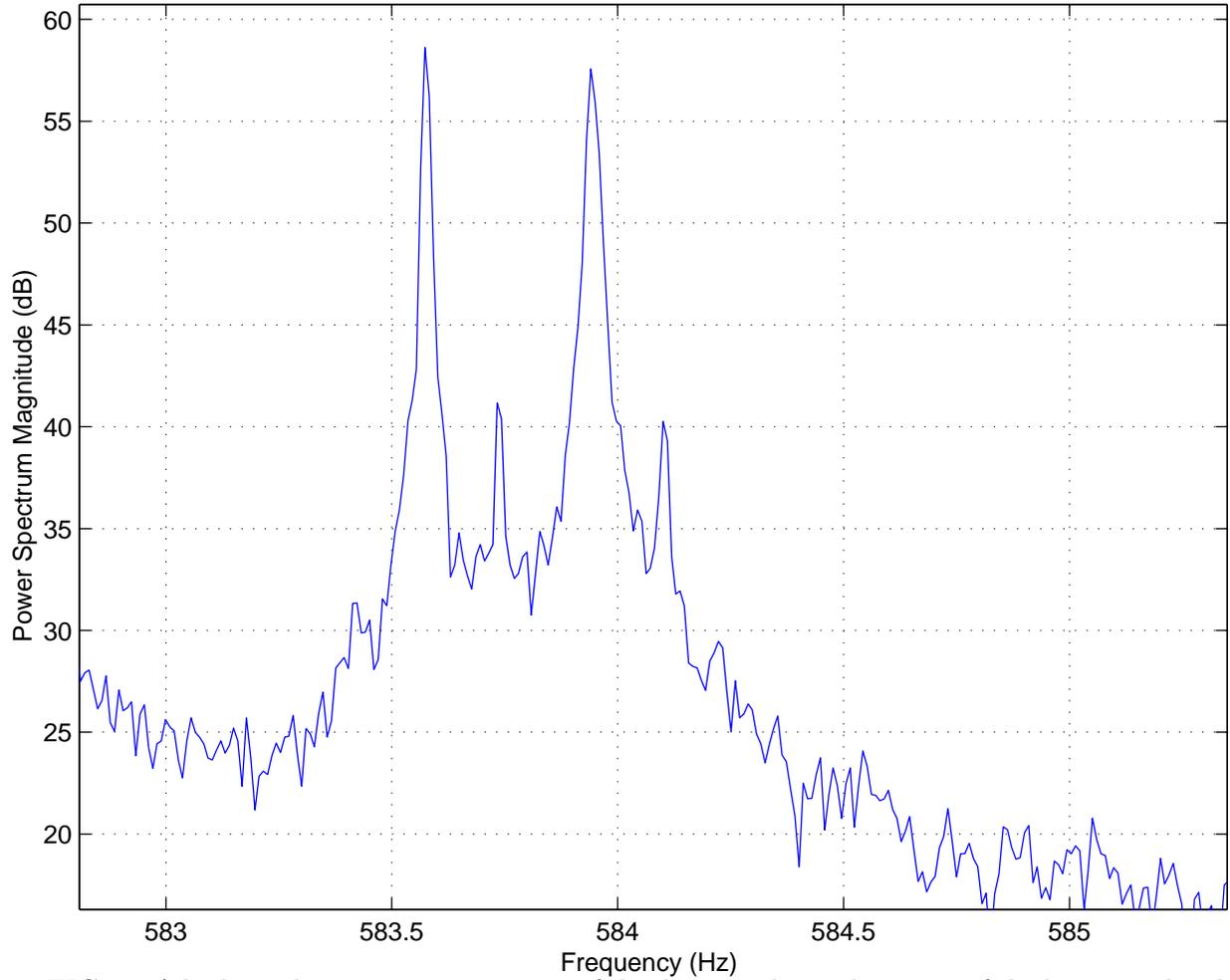}
\caption{A high resolution power spectrum of the data set shows that
  most of the large amplitude violin mode resonances have a lower
  amplitude satellite line at slightly higher frequency. Here we shows
  the high amplitude lines together with their satellites in the 583.3
  and 584.3 Hz band. For more detail see section
  \protect\ref{sec:genchar}.} \label{fig:sat1}
\end{figure}

Referring to table \ref{tbl:lines} note that, apart from modes 7, 16,
19, 24, and 25, the lines appear in pairs, with a high amplitude line
closely associated with a lower amplitude line at a slightly higher
frequency (less than or of order 0.2~Hz). This association of lines is
clearly shown in figure~\ref{fig:sat1}. Of the five apparently
unpaired lines, three show companions in a high resolution power
spectrum taken in the band of the first harmonic, suggesting that all
five of the ``missing modes'' may be so closely spaced in frequency
with other, identified modes, that we are unable to resolve them.

\subsection{Application}\label{sec:subapp}

\begin{figure}
\epsfxsize=\columnwidth 
\epsffile{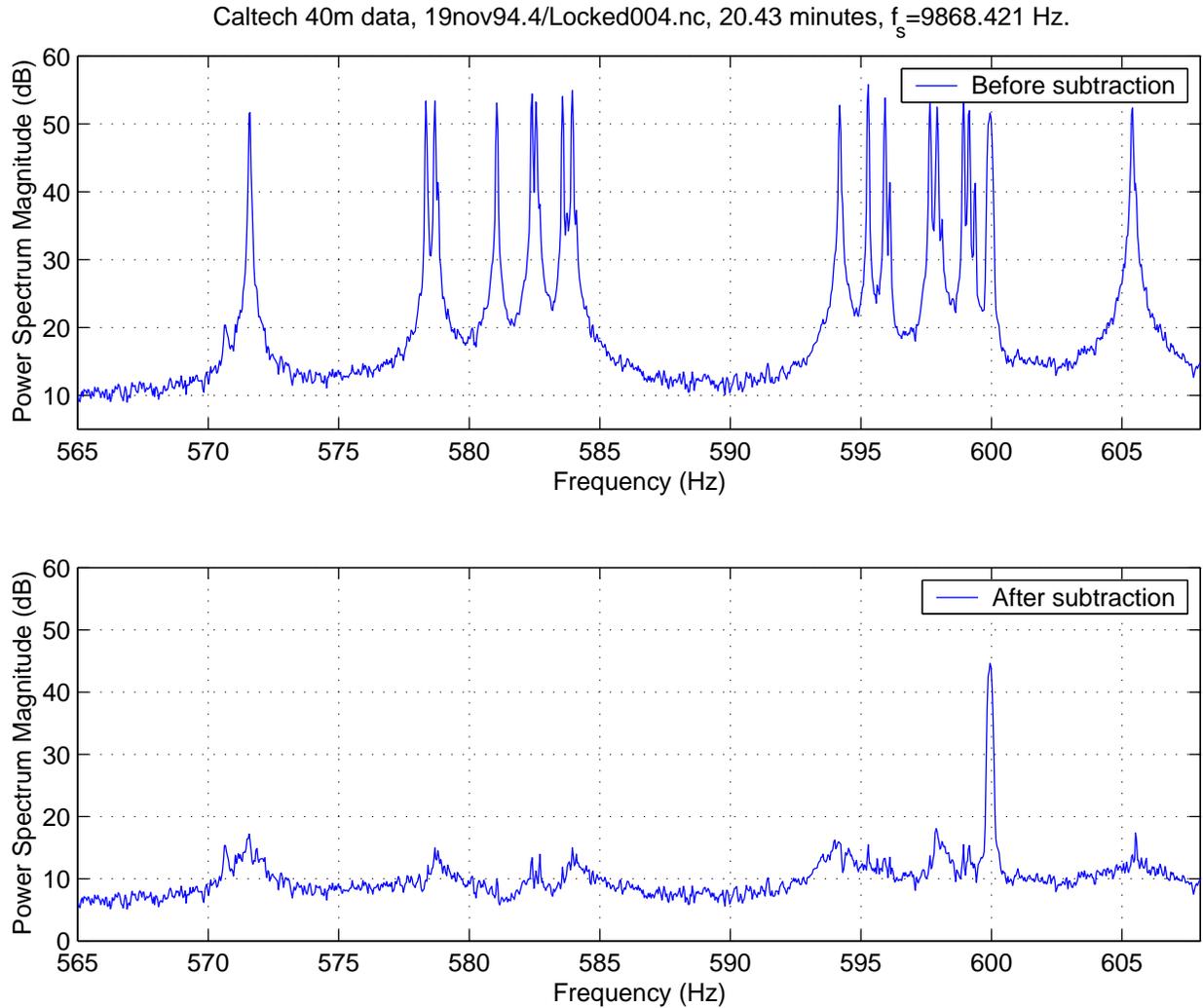}
\caption{Top panel: A high resolution power spectrum of the LIGO 40~M
  prototype data, focusing on the 571 to 605 Hz band, shows the violin
  modes standing 40 to 50~dB above the noise floor. Bottom panel: A
  power spectrum of the same data, after the Kalman filter prediction
  for the mode contribution to the measurement has been subtractively
  removed, shows that the modes have been removed, leaving only a
  small residual in addition to the measurement noise. For more
  details see section \protect\ref{sec:subapp}.}\label{fig:40power}
\end{figure}

Using the techniques described in section \ref{sec:model} we have
applied a Kalman filter designed to track the 27 violin modes
identified in table \ref{tbl:lines}. The top panel of
figure~\ref{fig:40power} shows the power spectrum of the {\tt
  IFO\_DMRO\/} (interferometer differential mode read-out, or
``gravity wave'') channel in the 45~Hz band beginning at 565.0 Hz,
with the violin modes clearly present. The bottom panel shows the
power spectrum of the residual after the modes identified by the
filter have been subtractively removed. Very little power associated
with these modes remains.\footnote{The 600~Hz line, which is the ninth
  harmonic of the power-main, was not modeled in our analysis.}  

Figure \ref{fig:40power} also shows an overall approximately 2~dB
reduction in the background noise level within the band, which is due
to the ``wings'' of the predicted contribution of the violin modes to
the total noise. Our experience with simulated data (cf.\ sec.\ 
\ref{sec:case1}--\ref{sec:case4}) shows that the filter generally
makes fractional prediction errors whose magnitude is on order 1\% and
that, as we have implemented the filter, this error tends to
systematically over-estimate the contribution of the mode wings to the
total noise. This is consistent with spectrum of the residual found
here.

\subsection{Preliminary characterization of mode motion and
  residual}\label{sec:prelimChar} 

The Kalman filter provides us with a decomposition of the observation
into the contributions from the individual violin modes and a
residual, which is, in a least-squares sense, as free from the
influence of the modes as we can make it. The noise character of the
violin modes, which are instrumental artifacts, can thus be studied
separately from the noise character of the residual.

\begin{figure}
  \epsfxsize=\columnwidth \epsffile{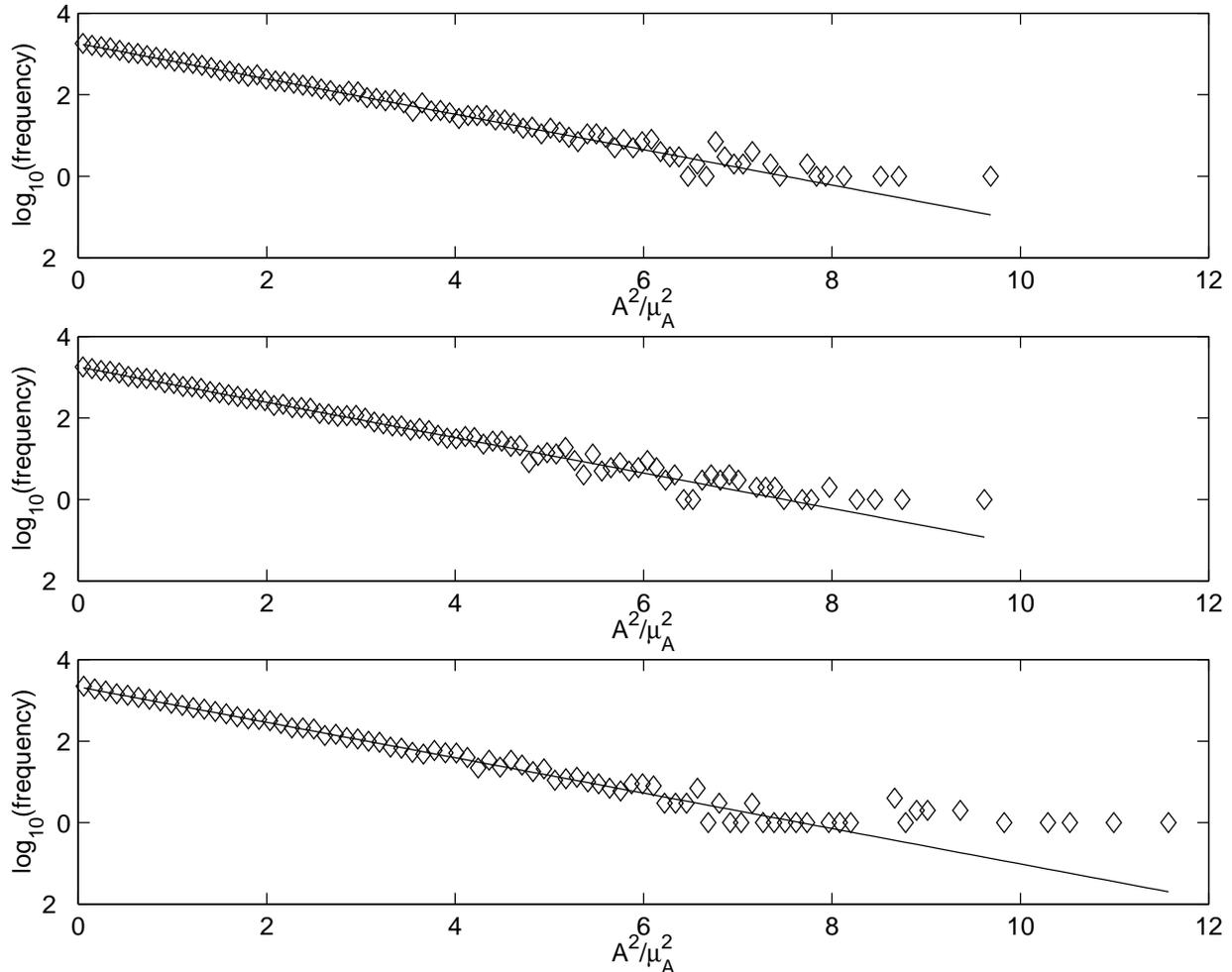} \caption{Histograms of
  squared-amplitudes in units of the amplitude variance.  The top
  panel shows the ``raw'' data from the {\tt IFO\_DMRO\/} channel in
  the 570--595~Hz band.  The middle panel histograms the amplitude of
  the Kalman estimate of the violin mode contributions to {\tt
  IFO\_DMRO}, while the bottom panel histograms the amplitude of the
  difference between the raw data and the Kalman estimate of the mode
  contributions.  In all cases the straight line is the expected
  result for Gaussian noise.  The raw data ``breaks'' at an
  amplitude-squared S/N of approximately 6 (which corresponds to the
  level at which single events are expected in each bin), after which
  it shows clearly an excess of high amplitude events.  The Kalman
  estimate of the contribution by the violin mode lines shows a very
  similar character, supporting the conclusion that the violin modes
  are dominating the noise in this band.  The residual, on the other
  hand, appears to show a modest break at an amplitude-squared S/N of
  4 and then a dramatically larger number of high amplitude events
  above a S/N of 6.  For more discussion see section
  \protect\ref{sec:prelimChar}.}\label{fig:40expstat}
\end{figure}

Figure~\ref{fig:40expstat} shows in histogram form the frequency with
which samples of given amplitude (normalized by rms amplitude) appear
in the $(570,595)\,\text{Hz}$ bandwidth of the original data
(``raw''), in the contribution to this band from the modes as
identified by the Kalman filter, and in the residual difference
between the raw data stream and the estimated contribution from the
modes.  If the detector noise were strictly Gaussian the distribution
of squared-amplitudes (normalized by the mean-squared amplitude) would
follow an exponential distribution and the resulting curves would be,
on the semi-log graph as shown, a straight lines with slope
$-\log_{10}e$.  We have evaluated $\chi^2$ for each histogram, using
the left-most 47 bins shown in figure \ref{fig:40expstat} and with the
remainder of the data bins collapsed into a single additional bin, so
that no bin has fewer than 40 events.  The same binning was used for
all three $\chi^2$ statistic evaluations.  Table \ref{tbl:chi2} gives
the resulting $\chi^2$, the number of degrees of freedom, and the
corresponding $p$-value.

Referring to table \ref{tbl:chi2} we see that the Kalman filter has
resolved the total noise in this band into two components that have
very different noise character.  While in all cases the statistical
character of the noise is far from Gaussian, the contribution
identified as arising from the violin modes is much better behaved
({\em i.e.,} much closer to Gaussian) than the residual.  The close
correspondence between the $\chi^2$ for the ``raw'' data and the
identified violin mode contribution reflects the fact that the violin
mode contribution, even while it is narrow-band, is of such great
amplitude that it dominates the total detector noise in this band. 
Recalling our earlier results (cf.\ sec.\ \ref{sec:case3} and
\ref{sec:case4}), which show how the Kalman filter is able to
discriminate correctly between excitations of the modes and impulses
in the measurement noise, we conclude that the difference in the noise
character is real and that, in the LIGO~40~M, the violin mode
contribution to the overall noise has many fewer outliers than the
non-mode background. Since, in a real detector, it is the residual 
that will contain the gravitational wave signal, the importance of 
removing instrumental artifacts like the violin modes in order to 
discover the noise character of the residual cannot be overstated. 

\begin{figure}
\epsfxsize=\columnwidth
\epsffile{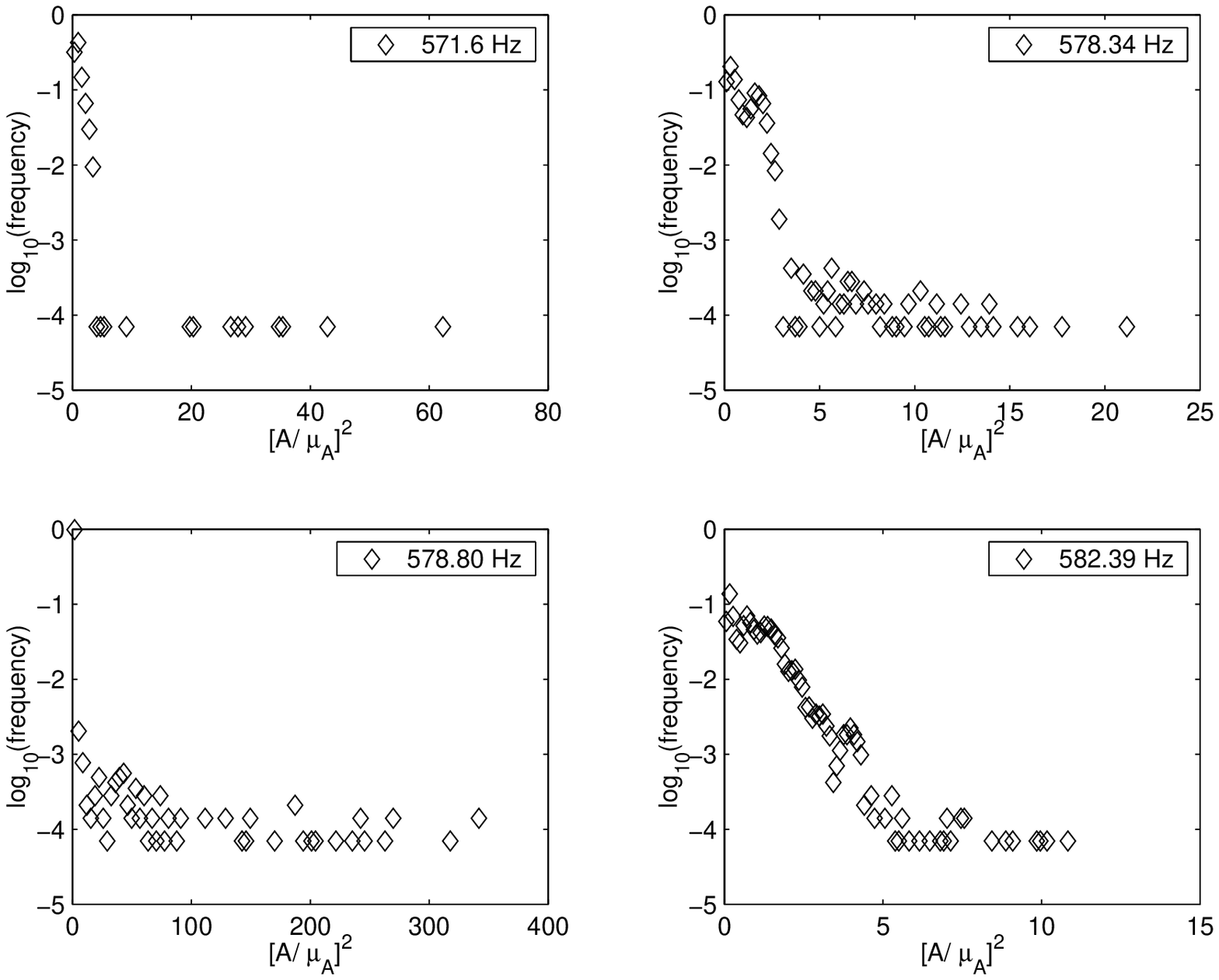}
\caption{Histograms of the square of Kalman estimated amplitude
  divided by mean amplitude for each of the individual Kalman
  estimated violin modes. For more discussion see section
\protect\ref{sec:prelimChar}.}\label{fig:indstat1} 
\end{figure}

\begin{figure}
\epsfxsize=\columnwidth
\epsffile{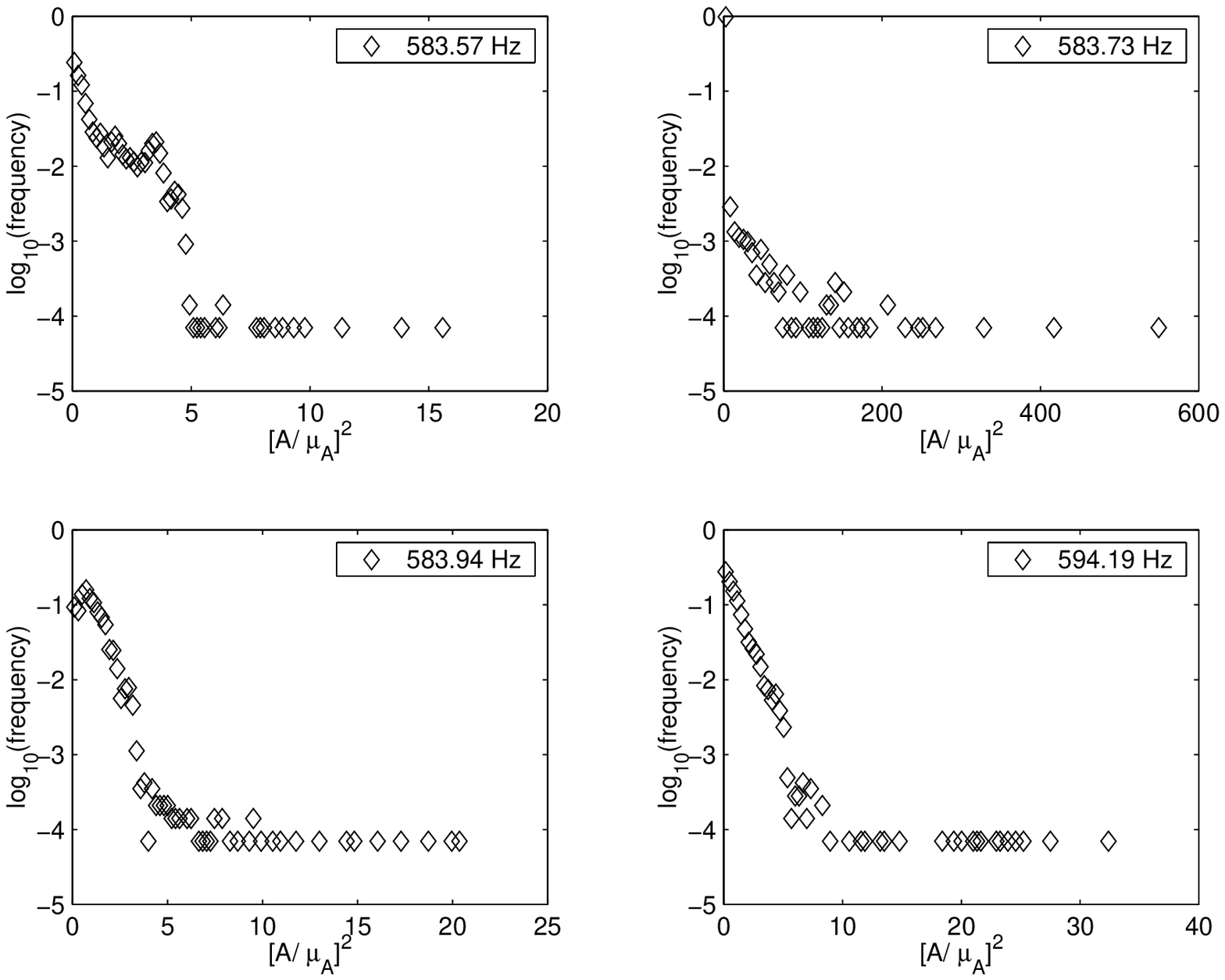}
\caption{Histograms of the square of Kalman estimated amplitude
divided by mean amplitude for each of the individual Kalman estimated
violin modes. For more discussion see section
\protect\ref{sec:prelimChar}.}\label{fig:indstat2}
\end{figure}

\begin{figure}
\epsfxsize=\columnwidth
\epsffile{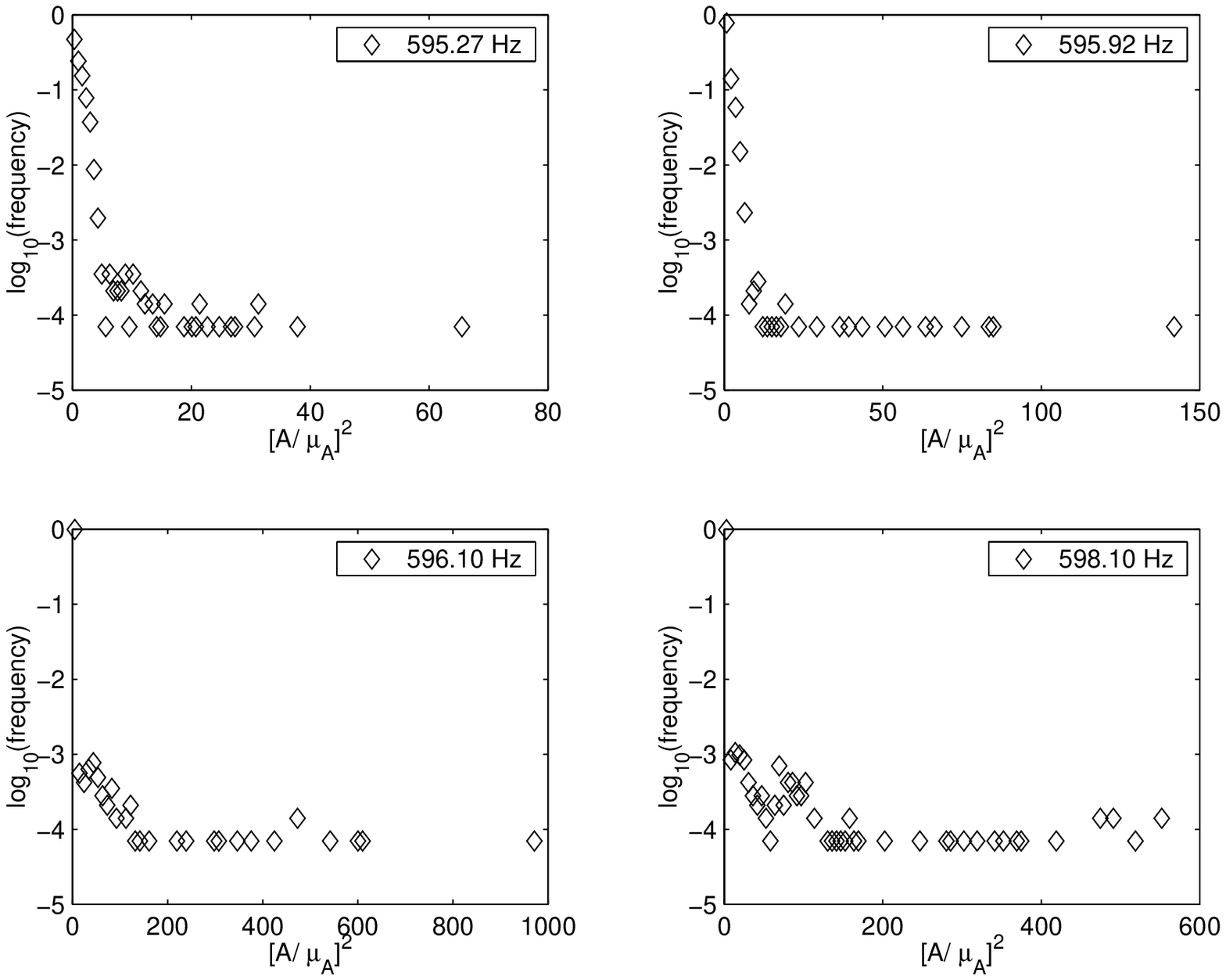}
\caption{Histograms of the square of Kalman estimated amplitude
divided by mean amplitude for each of the individual Kalman estimated
violin modes. For more discussion see section
\protect\ref{sec:prelimChar}.}\label{fig:indstat3}
\end{figure}

A still closer examination, focusing on the contributions from the
individual modes, shows that certain modes have much greater excess
noise than others. Figure~\ref{fig:indstat1}--\ref{fig:indstat3}
shows, for twelve different violin modes, the frequency distribution
of mode amplitudes.  All of the modes shown here have a strong excess
of high amplitude noise events: {\em i.e.,} they are subject to
strong, non-thermal forces.

\section{Conclusions}\label{sec:conclusions}

Violin modes of the test mass suspensions are a significant
instrumental artifact in the signal band of modern interferometric
gravitational wave detectors: at the mode resonant frequencies they
stand between 40 and 50~dB above the ambient noise floor.  These modes
are not excited or otherwise disturbed by the passage of a
gravitational wave; yet, they contribute almost all of the noise power
in a wide bandwidth.  Correspondingly, the ability to identify and
remove their contribution to the detector noise and characterize the
residual noise can significantly increase the detector's sensitivity
in the affected wave-band.

Treating the modes as a stochastic signal in a measurement noise
background we have developed and demonstrated the use of a Kalman
filter for identifying the state of the suspension wires and the
contribution of the violin modes to the detector noise.  A Kalman
filter uses the known dynamics of the process that underlies a noise
component (in this case the stochastic excitation of the suspension
violin modes) to identify that component's contribution to total
system noise from measurements made on the total noise alone.  When
the stochastic driving force is Normal ({\em i.e.,} Gaussian) the
Kalman filter is optimal in the least squares sense.  When the process
noise is non-Gaussian, the Kalman filter estimate is optimal when only
the second order statistics of the noise ({\em i.e.,} its power
spectrum) is known.

Using simulated data, whose mode and measurement noise contribution is
known exactly, we have shown that the filter works well in the
presence of both Gaussian and non-Gaussian measurement noise and
excitations of the violin modes. We have also shown that the filter
easily discriminates between transients in the measurement noise and
transient excitations of the violin modes themselves, capturing the
latter in its estimate of the mode state while rejecting the former as
not arising from excitation of the mode. Using data taken at the
LIGO~40~M we have demonstrated the use of the filter by isolating and
subtractively removing the noise owing to 27 distinct violin modes,
each standing between 40 and 50~dB above the ambient noise floor, over
an approximately 40~Hz bandwidth.

The Kalman filter describes here operates in the time domain,
providing instantaneous estimates of the suspension violin modes
states. The computational cost is negligible, making it suitable for
use as an on-line diagnostic as well as for on- or off-line data
analysis. 

Thermal and other excitations of the violin modes dominate the
detector noise over a wide band.  This additive contribution to the
detector output will never include any gravitational wave signal,
however.  Separating the mode contribution from the total noise
dramatically increases the detector's sensitivity by significantly
reducing the mean-square noise amplitude.  It also, however, reveals a
new ``layer'' of non-Gaussian and transient noise that must be
characterized and, where possible, removed through its association
with other instrumental and physical environment monitors.  Using the
Kalman filter to separate the total noise into a violin mode
contribution and a residual we find that, in the LIGO~40~M prototype,
the noise owing to the violin modes is much ``cleaner'' ({\em i.e.,}
much closer to Normal) than the residual noise, which is strongly
leptokurtic.  Since, in a real detector, it is the residual that will
contain the gravitational wave signal, the importance of removing
instrumental artifacts like the violin modes in order to discover the
noise character of the residual cannot be overstated.  As we dig
deeper into the detector noise the challenges of identifying the
origin of the newly uncovered noise components becomes greater;
however, this is a challenge that must be faced and overcome if we are
to make the most of the opportunities provided by the these new
detectors.

\section{Acknowledgment}

We are grateful to Albert Lazzarini for drawing our attention to
Kalman Filtering and suggesting that it might be a useful tool for
gravitational wave data analysis. We are also glad to thank the LIGO
Laboratory at the California Institute of Technology for their
hospitality during the 1997/8 academic year, when this work was
begun. We also gratefully acknowledge the use of data taken at the LIGO
40~M interferometer during November 1994.  SM wishes to thank Soumya D.
Mohanty for many valuable insights and criticisms, and Constantin Brif
for an extremely valuable set of notes on thermal noise in test mass
suspensions of interferometric gravitational wave detectors. LSF is
particularly glad to thank Peter Fritschel for valuable
discussions. Finally, we are glad to thank Gabriela Gonzalez, Gary
Sanders and David Tanner for their comments on this manuscript. 

This work was supported by National Science Foundation
award PHY 99-96213, PHY 98-00111 and their predecessors. 


\begin{thebibliography}{10}

\bibitem{hough96a}
J. Hough,  in {\em Proceedings of the 7th {M}arcel {G}rossman {M}eeting},
  edited by R.~T. Jantzen and G.~M. Keiser (World Scientific, Singapore, 1996).

\bibitem{abramovici92a}
A. Abramovici {\it et~al.}, Science {\bf 256},  325  (1992).

\bibitem{bradaschia90a}
C. Bradaschia {\it et~al.}, Nucl. Instrum. Methods Phys. Research {\bf A289},
  518  (1990).

\bibitem{kuroda97a}
K. Kuroda,  in {\em Gravitational {W}aves: {S}ources and {D}etectors}, edited
  by I. Ciufolini and F. Fidecaro (World Scientific, Singapore, 1997).

\bibitem{kalman60a}
R. Kalman, Trans. ASME, J. Basic Eng. {\bf 82D},    (1960).

\bibitem{sintes98a}
A. Sintes and B. Schutz, Phys. Rev. D {\bf 58},  122003  (1998).

\bibitem{sintes99a}
A.~M. Sintes and B.~F. Schutz, Phys. Rev. D {\bf 60},  062001  (1999).

\bibitem{allen98a}
B. Allen, GRASP: a data analysis package for gravitational wave detection,
  1998, available from
  $<$http://www.lsc-group.phys.uwm.edu/$\sim$ballen/grasp-distribution$>$.

\bibitem{lees95a}
J.~M. Lees and J. Park, Computers \& Geosciences {\bf 21},  199  (1995).

\bibitem{mukherjee99a}
S. Mukherjee and L.~S. Finn, Removing Instrumental Artifacts: Suspension Violin
  Modes, gr-qc/9911098, 1999, to appear in the (refereed) Proceedings of the
  Third Edoardo Amaldi Conference on the Detection of Gravitational Waves.

\bibitem{doob55a}
J.~L. Doob, {\em Stochastic Processes} (John Wiley \& Sons, New York, 1955).

\bibitem{brockwell87a}
P.~J. Brockwell and R.~A. Davis, {\em Time Series: Theory and Methods}
  (Springer-Verlag, New York, 1987).

\bibitem{gersch93a}
W. Gersch,  in {\em New Directions in Time Series Analysis, part II}, edited by
  A. Friedman and W.~M. Jr. (Springer-Verlag, New York, 1993).

\bibitem{brown97a}
R.~G. Brown and P.~Y.~C. Hwang, {\em Introduction to Random Signals and Applied
  Kalman Filtering} (John Wiley and Sons, New York, 1997).

\bibitem{gabypc}
G. Gonzalez, 2000, private communication.

\bibitem{strang96a}
G. Strang and T. Nguyen, {\em Wavelets and Filter Banks} (Wellesley-Cambridge
  Press, Wellesley, Massachusetts, 1996).

\bibitem{callen51a}
H.~B. Callen and T.~A. Welton, Phys. Rev. {\bf 83},  34  (1951).

\bibitem{matlab}
Matlab, a technical computing environment for high-performance numeric
  computations in linear algebra, is a product of {T}he {M}ath{W}orks, Inc.

\bibitem{gillespie95a}
A.~D. Gillespie, Ph.D. thesis, California Institute of Technology, 1995.

\end{thebibliography}

 
\begin{table} 
\begin{tabular}{c|ccccc} 
&Case 1&Case 2&Case 3&Case 4&\\ 
\hline 
$f_0\,(Q_0)$&571.6~Hz ($57\times10^{3}$)&571.6~Hz ($57\times10^{3}$)&571.6~Hz
($57\times10^{3}$)&571.6~Hz ($57\times10^{3}$)&\\ 
$f_1\,(Q_1)$&N/A&584.6~Hz($57\times10^{3}$)&N/A&N/A&\\ 
$f_2\,(Q_2)$&N/A&585.4~Hz($57\times10^{3}$)&N/A&N/A& 
\end{tabular} 
\caption{Constants characterizing the simulations described in 
section~\protect{\ref{sec:obs}}. The sample rate in all cases was 
  9868.421~Hz.}\label{tbl:SimParams}
\end{table} 

\begin{table}
\begin{tabular}{cccccc}
  Line&$\alpha$&$\mu_1$&$\sigma_1$&$\mu_2$&$\sigma_2$\\
 (Hz)&&&&&\\
\hline
571.6&0.7&0&1&0&1.5\\
585.0&0.7&0&1&0&1.5\\
585.6&0.7&0&1&0&1.5
\end{tabular}
\caption{The mixture-Gaussian model parameters used to describe the
 non-Gaussian process and measurement noise used in 
 section~\protect\ref{sec:case2}.}\label{tbl:ex2}
\end{table}

\begin{table}
\begin{tabular}{ccccc} 
Line& Frequency& FWHM &Q&Amp.\\
ID \#&(Hz)     &(Hz/100)&($10^4$) &(dB)\\ \hline 
1&571.5869&1.26&4.55&58.0\\ 
2&571.6820&2.10&2.72&38.0\\ 
3&578.3400&1.01&5.73&68.0\\ 
4&578.4185&1.40&4.13&43.0\\ 
5&578.6820&2.00&2.89&56.0\\ 
6&578.8050&1.31&4.42&46.0\\ 
7&581.0550&2.00&4.42&57.0\\ 
8&582.3957&1.70&3.43&58.0\\ 
9&582.5586&2.60&2.24&57.0\\ 
10&583.5750&1.55&3.76&60.0\\      
11&583.7379&1.64&3.56&40.0\\      
12&583.9429&1.90&3.07&57.0\\ 
13&584.1077&2.20&2.66&38.0\\ 
14&594.1913&1.70&3.50&58.0\\ 
15&594.2902&2.20&2.70&42.5\\ 
16&595.2734&1.30&4.58&61.0\\ 
17&595.9235&1.10&5.42&60.0\\ 
18&596.1029&1.50&3.97&46.0\\ 
19&597.6456&1.80&3.32&58.5\\ 
20&597.9212&1.50&3.99&57.0\\ 
21&598.1042&1.40&4.60&40.0\\ 
22&598.9301&2.00&2.99&60.0\\ 
23&599.0271&1.30&4.61&36.0\\ 
24&599.1425&1.40&4.28&60.0\\ 
25&599.3713&1.18&5.08&47.5\\ 
26&605.3880&4.50&1.34&55.0\\ 
27&605.5310&1.20&5.05&45.0   
\end{tabular}
\caption{Violin mode frequencies, full-width at half maximum (FWHM),
  derived damping constants $Q$, and mode amplitudes for the 27 violin
  modes found observed {\tt IFO\_DMRO\/} data channel of the LIGO~40~M
  prototype.  The modes and their characteristics were determined from
  high resolution power spectra formed from the data sets
  themselves. There was no significant variation in the mode frequency
  or quality over the three days covered by the November 1994 data
  set.}\label{tbl:lines}  
\end{table}

\begin{table}
\begin{tabular}{lrrr}
&total&violin&residual\\
&noise&mode&\\
\hline
$\chi^2$&39.8514&39.7136&47.1459\\
$N_{\text{freedom}}$&47&47&47\\
$p$-value&$0.7607$&$0.7656$&0.4666
\end{tabular}
  \caption{The correspondence between the statistical character of the
  LIGO~40~M detector noise, in the band 570--595~Hz about the violin
  modes, and a Normal distribution.  The figure of merit used in the
  comparison is $\chi^2$.  The first column gives $\chi^2$ for the
  total noise, the second column for that part of the noise attributed
  to the violin mode and the third column for the difference between
  the total noise and the violin mode contribution.  The violin mode
  contribution dominates the total noise in this band; when it is
  removed, the residual is seen to be very poorly behaved.  For more
  detail see the discussion in section
  \ref{sec:prelimChar}.}\label{tbl:chi2}
\end{table} 

\end{document}